\documentclass[12pt]{article}

\RequirePackage[OT1]{fontenc}
\usepackage{natbib}
\RequirePackage[colorlinks,citecolor=blue,urlcolor=blue]{hyperref}
\usepackage[english]{babel}
\usepackage{amsthm}
\usepackage{amsmath}
\usepackage{amsfonts}
\usepackage{amssymb}
\usepackage{graphicx}
\usepackage{enumerate}
\usepackage[normalem]{ulem}
\usepackage[dvipsnames]{xcolor}
\usepackage{array}
\usepackage{bm,bbm}

\newcommand{\bs}{\boldsymbol}
\def\calS{{\mathcal{S}}}
\DeclareSymbolFont{AMSb}{U}{msb}{m}{n}
\DeclareMathSymbol{\Real}{\mathbin}{AMSb}{"52}
\def\pr{{\rm Pr}}
\def\Dto{{\ {\buildrel D\over \longrightarrow}\ }}
\def\E{{\rm E}}
\def\calP{{\mathcal{P}}}
\def\calD{{\mathcal D}}
\def\calO{{\mathcal{O}}}

\begin{document}
\thispagestyle{empty}
\baselineskip=28pt
\vskip 5mm
\begin{center} 
{\Large{\bf Advances in Statistical Modeling of Spatial Extremes}}
\end{center}

\baselineskip=12pt
\vskip 5mm

\begin{center}
\large
Rapha\"el Huser$^1$ and Jennifer L. Wadsworth$^2$
\end{center}

\footnotetext[1]{
\baselineskip=10pt Computer, Electrical and Mathematical Sciences and Engineering (CEMSE) Division, King Abdullah University of Science and Technology (KAUST), Thuwal 23955-6900, Saudi Arabia. E-mail: raphael.huser@kaust.edu.sa}
\footnotetext[2]{
\baselineskip=10pt Department of Mathematics and Statistics, Fylde College, Lancaster University, Lancaster, United Kingdom. E-mail: j.wadsworth@lancaster.ac.uk}

\baselineskip=17pt
\vskip 4mm
\centerline{\today}
\vskip 6mm

\begin{center}
{\large{\bf Abstract}}
\end{center}
The classical modeling of spatial extremes relies on asymptotic models (i.e., max-stable processes or $r$-Pareto processes) for block maxima or peaks over high thresholds, respectively. However, at finite levels, empirical evidence often suggests that such asymptotic models are too rigidly constrained, and that they do not adequately capture the frequent situation where more severe events tend to be spatially more localized. In other words, these asymptotic models have a strong tail dependence that persists at increasingly high levels, while data usually suggest that it should weaken instead. Another well-known limitation of classical spatial extremes models is that they are either computationally prohibitive to fit in high dimensions, or they need to be fitted using less efficient techniques. In this review paper, we describe recent progress in the modeling and inference for spatial extremes, focusing on new models that have more flexible tail structures that can bridge asymptotic dependence classes, and that are more easily amenable to likelihood-based inference for large datasets. In particular, we discuss various types of random scale constructions, as well as the conditional spatial extremes model, which have recently been getting increasing attention within the statistics of extremes community. We illustrate some of these new spatial models on two different environmental applications.
\baselineskip=16pt

\par\vfill\noindent

\pagenumbering{arabic}
\baselineskip=24pt

\newpage


\section{Introduction}\label{sec:Introduction}

\subsection{Motivation}\label{subsec:Motivation}
The spatial and/or temporal modeling of extreme events is fundamental for assessing risks in climatology \citep{Blanchet.Davison:2011,Davison.Gholamrezaee:2012,Gilleland.etal:2013,Stephenson.etal:2015,Schar:2016,Risser.Wehner:2017,Bopp.Shaby:2017,Reich.Shaby:2019}, hydrology \citep{Katz.etal:2002,Cooley.etal:2007,Thibaud.etal:2013,Huser.Davison:2014a,Bopp.etal:2020b,Bacro.etal:2020}, ecology \citep{Thibaud.etal:2016}, 
oceanography \citep{Jonathan.Ewans:2013,Huser.Wadsworth:2019,Shooter.etal:2019}, public health \citep{Eastoe.Tawn:2009,Vettori.etal:2019,Vettori.etal:2020}, finance and economics \citep{Smith.Katz:2013,CastroCamilo.etal:2018}, among other fields. 

Extreme-Value Theory (EVT) provides a natural, elegant, and mathematically rigorous framework for approaching this problem, modeling rare events and assessing such risks. Although several textbooks \citep{Embrechts.etal:1997,Coles:2001,Beirlant.etal:2004,deHaan.Ferreira:2006,Reiss.Thomas:2007} and review papers \citep{Davison.etal:2012,Cooley.etal:2012a,Segers:2012,Ribatet:2013b,Davison.Huser:2015,Cooley.etal:2019,Davison.etal:2019} have already been written on the classical theory and application of univariate, multivariate and spatial extremes, more recent topics that transcend the classical framework have not been covered in depth. In this review paper, we intend to fill this gap by providing a modern up-to-date account {of} recent advances in the spatial modeling of extreme events, focusing on flexible models and alternative formulations that allow bridging asymptotic dependence classes. 

This review paper may be read as a follow-up of \citet{Davison.etal:2012}, which is an excellent paper on the classical modeling of spatial extremes. \citet{Engelke.Ivanovs:2021} is another recent review that covers advances in sparse models for multivariate extremes, a topic of major interest nowadays. To be concise, we shall not treat this topic here. Moreover, as the univariate and multivariate modeling of extremes are already covered in depth in the literature, we refer to the aforementioned 
textbooks and review papers for more details on these topics. 

\subsection{Classical univariate and spatial extreme-value models}\label{subsec:ClassicalEVT}
In the univariate context, classical EVT relies on asymptotic extreme-value models for block maxima or high threshold exceedances. While the generalized extreme-value (GEV) distribution arises as the only possible limit model for block maxima, 
the generalized Pareto (GP) distribution is its counterpart for high threshold exceedances. 
Both limit distributions are intimately connected through a point process characterization \citep{Davison.Smith:1990}, and have been widely used for modeling extremes, either defined as block maxima or high threshold exceedances, respectively. Although the threshold exceedance approach is usually preferred nowadays over block maxima because it allows one to have a more detailed modeling of extremal clusters due to temporal dependence and to incorporate more data in estimation, the choice of one approach or the other is often dictated by the context. 
The extreme-value paradigm assumes that the limit GEV and GP distributions are good approximations for block maxima and threshold exceedances in finite samples, and that these models fitted at high but finite levels can be used for extrapolation beyond observed data. 


In the spatial context, the definition of an extreme event is less clear. One possibility is to model spatially-indexed block maxima using max-stable processes \citep{Padoan.etal:2010}, which are the natural generalization of the GEV distribution to the infinite-dimensional setting. These asymptotic models have received a lot of attention over the last decade and have been used in a wide variety of environmental applications; see, e.g., \citet{Padoan.etal:2010}, \citet{Blanchet.Davison:2011}, 
\citet{Reich.Shaby:2012a}, \citet{Opitz:2013a}, \citet{Stephenson.etal:2015} \citet{Huser.Genton:2016} and \citet{Oesting.etal:2017}. However, it is difficult to make inference for max-stable models in high-dimensional applications due to the complicated form of the associated likelihood function \citep{Padoan.etal:2010,Ribatet.etal:2012,Thibaud.etal:2016,Castruccio.etal:2016,Huser.etal:2016,Huser.etal:2019}, and the simulation and conditional simulation algorithms that are both expensive to run and tedious to implement \citep{Schlather:2002,Oesting.etal:2012,Dombry.etal:2013,Dieker.Mikosch:2015,Dombry.etal:2016,Liu.etal:2019}. Moreover, the block maximum approach has been criticized in the spatial context for relying on artificially created spatial block maxima and not directly modeling the actual observed spatial events that effectively took place.


Alternatively, a spatial process $Y(\bs{s})$, $\bs{s}\in\calS\subset\Real^d$, may be defined as extreme when a suitable scalar functional of $Y$ exceeds some high threshold. By analogy with multivariate generalized Pareto distributions, one possible choice is to consider conditioning on the event $\sup_{\bs{s} \in \mathcal{S}} Y(\bs{s})$ being large, which leads to generalized Pareto processes \citep{Ferreira.deHaan:2014}. Other conditioning events may also be considered, which can be described by certain types of \emph{risk functionals} $r(\cdot)$ applied to the process on a standardized scale. The limit models that arise under a suitable asymptotic regime for conditional $r$-threshold exceedances are called $r$-Pareto processes \citep{Dombry.Ribatet:2015,Thibaud.Opitz:2015,deFondeville.Davison:2018}. 
The benefits of the threshold exceedance approach in the spatial context are that the observed spatial processes are modeled directly, rather than artificially relying on pointwise maxima, and that the likelihood function is usually simpler compared to the block maximum approach based on max-stable processes. Nevertheless, there are still some computational difficulties depending on the choice of risk functional $r(\cdot)$, and full likelihood estimators are typically highly biased if spatial extreme events include marginally non-extreme observations. The bias problem may often be dealt with by censoring non-extreme observations, but this approach is computationally demanding in high dimensions due to the multi-fold integrals involved. To circumvent this computational issue, \citet{deFondeville.Davison:2018} proposed an efficient inference approach based on scoring rules, which mimics the effect of censoring, while avoiding the intensive computation of integrals and the density normalizing constant. However, such an approach has other drawbacks, such as being difficult to adapt to the Bayesian framework, where inference is commonly performed using Markov chain Monte Carlo algorithms which rely on the likelihood function.

\subsection{Recent developments and paper outline}\label{subsec:Recent}
The asymptotic characterization of max-stable models and $r$-Pareto models may be seen as an appealing justification to fit such models in practice. However, the max-stability or threshold-stability properties of these asymptotic models yield quite rigid dependence structures, which may not hold at finite levels and may negatively affect the estimation of spatial risk measures. A related limitation of max-stable and Pareto processes is that they are always asymptotically dependent, unless they are fully independent. To be more precise, a stochastic process $Y(\bs{s})$ defined over a region $\calS\subset\Real^d$ is said to be asymptotically dependent if for any two sites $\bs{s}_1,\bs{s}_2\in\calS$, the random variables $Y(\bs{s}_1)\sim F_1,Y(\bs{s}_2)\sim F_2$ (whose generalized inverse denoted by $F_1^{-1}$ and $F_2^{-1}$, respectively) are such that the conditional exceedance probability $\chi_u(\bs{s}_1,\bs{s}_2)=\pr\{Y(\bs{s}_1)>F_1^{-1}(u)\mid Y(\bs{s}_2)>F_2^{-1}(u)\}$ has a positive limit as $u\to1$, i.e., 
\begin{equation}\label{eq:chi}
\chi(\bs{s}_1,\bs{s}_2)=\lim_{u\to1} \chi_u(\bs{s}_1,\bs{s}_2)>0.
\end{equation}
It is said to be asymptotically independent if this limit is zero, i.e., $\chi(\bs{s}_1,\bs{s}_2)=0$ in \eqref{eq:chi}. An asymptotically dependent process $Y(\bs{s})$ (like max-stable and Pareto processes) is such that extreme events have a positive probability to occur simultaneously at distinct sites, no matter how extreme they are. In other words, the spatial dependence strength does not vanish as events become more extreme. In practice, however, environmental data often tend to exhibit weakening dependence (i.e., to be spatially more ``localized'') for increasing quantile levels and to support asymptotic independence, although the asymptotic dependence class is usually unclear. This has motivated the development of models for asymptotic independence and more flexible hybrid models that can bridge the two asymptotic dependence regimes, often fitted to peaks over high thresholds; see, e.g., \citet{Wadsworth.Tawn:2012b}, \citet{Davison.etal:2013}, \citet{Opitz:2016}, \citet{Wadsworth.etal:2017}, \citet{Huser.etal:2017}, \citet{Huser.Wadsworth:2019}, and \citet{Bacro.etal:2020}. As these spatial models are designed to accurately capture the joint tail decay rate at high but \emph{finite} levels rather than describing the limiting dependence structure of extreme events, they are often referred to as \emph{sub-asymptotic} models by contrast with the \emph{asymptotic} max-stable and $r$-Pareto processes. Analogous models designed for block maxima, which similarly extend asymptotic models while remaining in the ``neighborhood'' of some popular max-stable processes, were also recently proposed by \citet{Bopp.etal:2020} and \citet{Huser.etal:2020}. 

Alternatively, the conditional spatial extremes approach, which aims at describing the spatial behavior of a random process conditional on single points being large, has recently been introduced as an alternative modeling strategy \citep{Wadsworth.Tawn:2019,Shooter.etal:2019}. Beyond having an asymptotic justification, the great benefit of this model lies in its flexibility to capture a wide range of asymptotic dependence behaviors, including changes in the asymptotic dependence class as a function of the distance between sites. Moreover, the model can be easily and quickly fitted in reasonably large dimensions using a likelihood-based approach, which bypasses censoring of non-extreme observations. The disadvantage of this conditional approach is that the model is more difficult to interpret ``unconditionally'', and that there is usually no obvious candidate for the conditioning site in spatial applications, though \citet{Wadsworth.Tawn:2019} have proposed a solution consisting in combining likelihood contributions from all sites. 


The rest of the paper is organized as follows. In Section~\ref{sec:asymptotic}, we review classical extreme-value theory in the univariate and spatial contexts, and we describe asymptotic extreme-value models and their likelihood-based inference approaches. In Section~\ref{sec:AIAD}, we introduce several classes of recently proposed sub-asymptotic models for spatial extremes, which can bridge asymptotic dependence and independence. In Section~\ref{sec:Conditional}, we present the conditional spatial extremes model. In Section~\ref{sec:applications}, we illustrate some of these recently proposed spatial models for the modeling of high threshold exceedances in two different environmental applications. Finally, we conclude in Section~\ref{sec:conclusion} with some discussion and perspectives on future research.

\section{Asymptotic models for spatial extremes}\label{sec:asymptotic}

\subsection{Marginal modeling of extremes}\label{subsec:univariate}
\begin{figure}[t!]
\centering
 \includegraphics[width=0.8\textwidth]{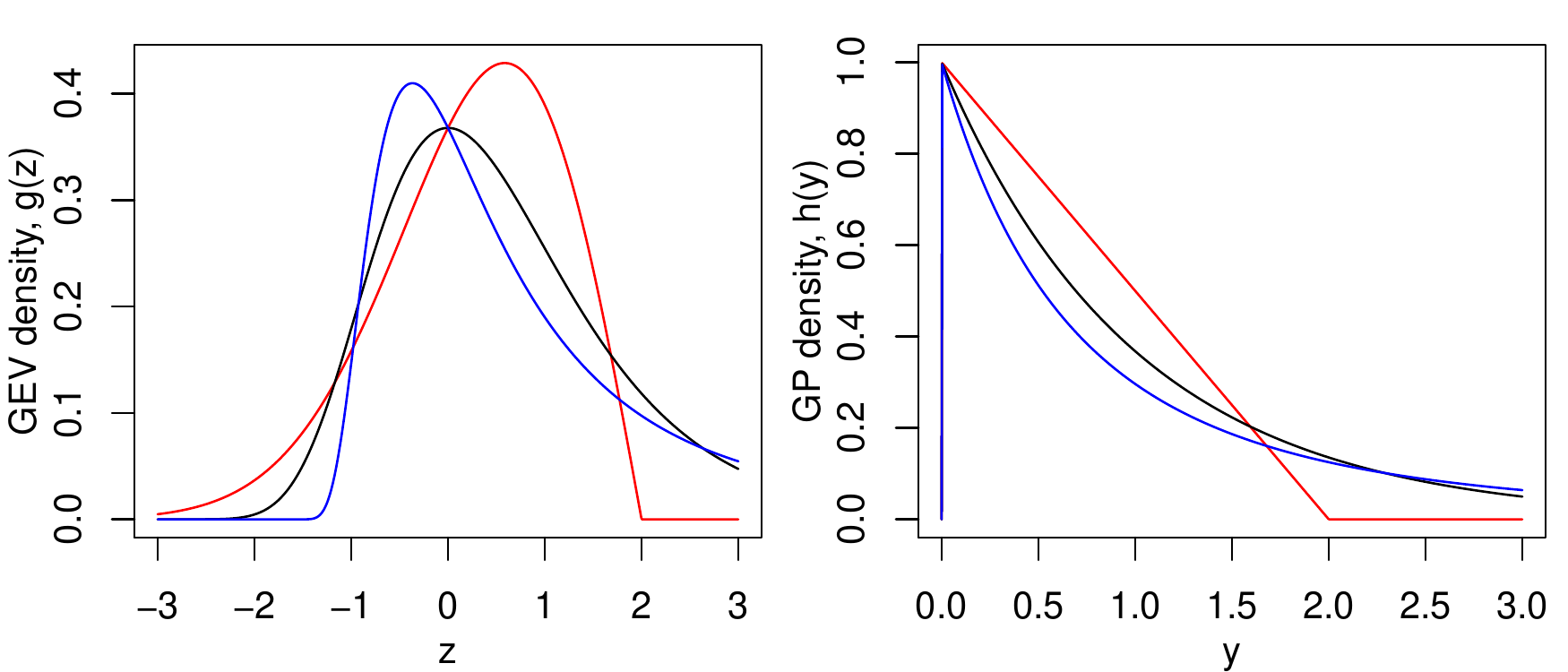}
  \caption{Left: Generalized extreme-value (GEV) density with parameters $\mu=0$, $\sigma=1$ and $\xi=-0.5$ (red), $\xi=0$ (black) and $\xi=0.5$ (blue). Right: Generalized Pareto (GP) density with parameters $\tau=1$ and $\xi=-0.5$ (red), $\xi=0$ (black) and $\xi=0.5$ (blue).}
 \label{fig:gevgp}
\end{figure}

The univariate theory of extremes is well understood, and its use in applications is by now quite standard. We recall key details here; see \citet{Coles:2001} for an introductory exposition. Let $Y_1,Y_2,\ldots$ be a sequence of independent random variables with common distribution $F$ and upper endpoint $y_F=\sup\{y\in\Real: F(y)<1\}$, and let $M_n=\max(Y_1,\ldots,Y_n)\sim F^n$. If there exist sequences of constants $a_n>0$ and $b_n\in\Real$ such that, as $n\to\infty$, 
\begin{equation}\label{eq:limitmax}
{M_n-b_n\over a_n}\Dto Z\sim G,
\end{equation}
where $G$ is a non-degenerate distribution and $\Dto$ denotes convergence in distribution, then the limit $G$ may be expressed as $G(z)=\lim_{n\to\infty}F^n(a_n z+b_n)$ and is necessarily of the form
\begin{equation}\label{eq:GEV}
G(z)=\left\{\begin{array}{ll}
\exp\left[-\left\{1+\xi(z-\mu)/\sigma\right\}_+^{-1/\xi}\right],&\xi\neq0,\\
\exp\left[-\exp\{-(z-\mu)/\sigma\}\right],&\xi=0,
\end{array}\right.
\end{equation}
for some parameters $\mu\in\Real$ (location), $\sigma>0$ (scale) and $\xi\in\Real$ (shape), where $a_+=\max(0,a)$ and with support $S_G=\{z\in\Real: 1+\xi(z-\mu)/\sigma>0\}$. The distribution \eqref{eq:GEV}, called the generalized extreme-value (GEV) distribution, is illustrated in Figure~\ref{fig:gevgp}. 
The shape parameter $\xi$, also called tail index, characterizes the tail behavior of $G$ with a bounded upper tail when $\xi<0$, a light tail with $\xi=0$, and a heavy tail when $\xi>0$. A key property of the limit distribution $G$ is that of \emph{max-stability}, which means that for any positive real $t>0$, there exist scalars $\alpha_t>0$ and $\beta_t\in\Real$ such that for all $z$,
\begin{equation}\label{eq:maxstability}
G^t(\alpha_t z+ \beta_t)=G(z).
\end{equation}
This implies that the distributions $G$ and $G^t$ are of the same type, i.e., they belong to the same location-scale family. In other words, $G$ and $G^t$ are both GEV distributions with the same tail index $\xi$, but different location and scale parameters. This property may be exploited for extrapolation beyond the observed data. To be concrete, assume that $Z_1,\ldots,Z_N$ represent independent yearly maxima of some variable of interest $Y$, say daily precipitation, measured over $N$ years at some fixed location, and assume that these maxima are modeled using a GEV distribution $G(z)$. Because the GEV distribution is max-stable, the distribution of the maximum over $k$ years (with $k$ potentially much larger than $N$), i.e., $\max(Z_1,\ldots,Z_k)$, is $G\{(z-\beta_k)/\alpha_k\}$ for some constants $\alpha_k>0$ and $\beta_k$ that can be estimated from the yearly maxima; thus, this relation can be used to rigorously estimate high quantiles that lie far into the upper tail, beyond the observed maximum. 

Assuming \eqref{eq:limitmax} holds, then high threshold exceedances $Y-u\mid Y>u$ may be approximated by the generalized Pareto (GP) distribution, in the sense that there exists a scaling function $a(u)>0$ such that 
\begin{equation}\label{eq:limitthr}
\pr\left({Y-u\over a(u)}>y\mid Y>u\right)\to 1-H(y):=\left\{
\begin{array}{ll}
(1+\xi y/\tau)_+^{-1/\xi}, & \xi\neq0,\\
\exp(-y/\tau), & \xi=0,
\end{array}\right.\quad u\to y_F,
\end{equation}
where $\tau>0$ and $\xi\in\Real$ are scale and shape parameters and the limit distribution $H$ in \eqref{eq:limitthr} is the GP distribution with parameters $\tau$ and $\xi$; see Figure~\ref{fig:gevgp}. The shape parameters $\xi$ in \eqref{eq:limitmax} and \eqref{eq:limitthr} are equal, but the scale parameters $\sigma$ and $\tau$ are different in general. By analogy with the GEV distribution being max-stable, the GP distribution is threshold-stable. We can show that if $Y-u\mid Y>u$ follows the GP distribution with parameters $\tau$ and $\xi$, then for all thresholds $v>u$, $Y-v\mid Y>v$ also follows the GP distribution with scale parameter $\tau+\xi(v-u)$ and shape parameter $\xi$. In other words, exceedances over higher thresholds remain GP with a modified scale parameter but with the same shape parameter $\xi$.

While the block maximum approach based on the GEV distribution and the threshold exceedance approach based on the GP distribution may seem quite different at first sight, they can be unified through a point process representation. Assume that \eqref{eq:limitmax} holds, and consider the bivariate point process $P_n=\{({i\over n+1},{Y_i-b_n\over a_n}); i=1,\ldots,n\}$ of rescaled event times and renormalized observations, respectively, where the sequences $a_n>0$ and $b_n$ are the same as in \eqref{eq:limitmax}, stabilizing the behavior of block maxima. Then, on regions of the form $A=[t_1,t_2]\times[u,+\infty]$, with $u$ bounded away from the lower endpoint of the limit GEV distribution $G$, the point process $P_n$ converges to a Poisson point process with mean measure
\begin{equation}\label{eq:limitPP}
\Lambda([s_1,s_2]\times[y,+\infty]) = (t_2-t_1)\{1+\xi (y-\mu)/\sigma\}_+^{-1/\xi},\quad {t_1\leq s_1<s_2\leq t_2, y>u.}
\end{equation}
The result for block maxima \eqref{eq:limitmax} can be obtained from \eqref{eq:limitPP} thanks to the void set probabilities of the limiting Poisson process, noting that the event $\{M_n\leq z\}$ is equivalent to having no points of $P_n$ in the set $A_z=[0,1]\times[z,+\infty]$. Similarly, the result for threshold exceedances \eqref{eq:limitthr} can be obtained by setting $u\equiv u_n = a_n u^\star + b_n$ for some fixed value $u^\star$, which tends to $y_F$ as $n\to\infty$, and $a(u)\equiv a(u_n)= a_n$, and noting that the left-hand side of \eqref{eq:limitthr} is equal to 
\begin{align*}
{\pr(Y>a(u)y+u)\over\pr(Y>u)}&\to {\Lambda([0,1]\times[y+u^\star,+\infty])\over \Lambda([0,1]\times[u^\star,+\infty])},\qquad n\to\infty\\
&= {\{1+\xi (y+u^\star-\mu)/\sigma\}_+^{-1/\xi}\over \{1+\xi (u^\star-\mu)/\sigma\}_+^{-1/\xi}},\qquad \xi\neq0\\
&= (1 + \xi y/\tau)_+^{-1/\xi},
\end{align*}
where $\tau=\sigma+\xi(u^\star-\mu)$. When $\xi=0$, the expression is obtained as the limit as $\xi\to0$. The point process characterization may also be used to build a statistical model for the $r$-largest order statistics, with $r=1,2,\ldots$, which extends the block maximum approach based on the GEV distribution, which arises for $r=1$.

In practice, unknown parameters may be estimated using a variety of techniques, and maximum likelihood and Bayesian inference approaches are particularly convenient because of their appealing large sample properties and their flexibility to handle complex settings. The selection of sample extremes depends on each approach (either based on block maxima, the $r$-largest order statistics, or threshold exceedances), and each limiting characterization has its own likelihood function formulation. In simple settings, a direct numerical optimization may be used to maximize the likelihood function and obtain marginal parameter estimates. In the non-stationary context, a generalized additive modeling (GAM) approach may be used by including covariates or splines into model parameters, and a penalized likelihood approach may be employed for inference \citep{ChavezDemoulin.Davison:2005,Northrop.Jonathan:2011,Jonathan.Ewans:2013,Jonathan.etal:2014}.

In the spatial setting, assume that $Y_1(\bs{s}),Y_2(\bs{s}),\ldots$ denotes a sequence of independent random processes defined over the region $\calS\subset\Real^d$, and observed at a finite collection of sites $\bs{s}_1,\ldots,\bs{s}_D\in\calS$. While the extremes within the time series $Y_i(\bs{s}_j)$, $i=1,2,\ldots$, could be analyzed separately at each site $\bs{s}_j$, it may be more sensible to model them jointly in a single model that links the data at the different sites together. One reason is that extreme events are sparse by definition, and spatial modeling allows to borrow strength across locations for better marginal estimation. In particular, the tail index $\xi$ can usually be assumed to be constant (or to vary smoothly) over an entire spatial region, and this dramatically reduces its estimation uncertainty, thereby also improving the subsequent estimation of high quantiles. Another reason is that prediction at unobserved locations may be required, and this can only be achieved with a proper spatial model. To this aim, Bayesian hierarchical models with a Gaussian latent structure are particularly convenient; see, e.g., \citet{Casson.Coles.1999}, \citet{Cooley.etal:2007}, \citet{Sang.Gelfand:2009}, \citet{Cooley.Sain:2010}, \citet{Turkman.etal:2010}, \citet{Davison.etal:2012}, \citet{Dyrrdal.etal:2015}, \citet{Geirsson.etal:2015}, \citet{Opitz.etal:2018} and \citet{CastroCamilo.etal:2019}. 
Such models, which are the Bayesian analogues of GAMs, can easily handle non-stationarity by embedding covariates into model parameters, as well as different types of latent Gaussian random effects that may be correlated over space and time---often specified with a sparse precision (i.e., inverse covariance) matrix to speed up computations. The smoothness of random effects can be regulated through a careful choice of prior distributions. They can be fitted using simulation-based Markov chain Monte Carlo (MCMC) methods, possibly in two steps by exploiting Max-and-Smooth \citep{Hrafnkelsson.etal:2020}, which relies on a Gaussian approximation to the likelihood, or by efficiently taking advantage of astute numerical techniques such as the integrated nested Laplace approximation (INLA; see \citealp{Rue.etal:2009}, \citealp{Rue.etal:2017}, \citealp{Bakka.etal:2018}, \citealp{Opitz.2017} and \citealp{Opitz.etal:2018}). However, all the above-mentioned latent Gaussian models assume that sample extremes are mutually independent conditional on some latent, spatially structured random effects included in model parameters. This conditional independence assumption, made for computational convenience, is often not realistic and may lead to a drastic underestimation of the joint occurrence of extreme events. 

To assess the joint behavior of spatial extreme events and accurately estimate their co-occurrence probabilities, more specialized models are required. {\citet{Sang.Gelfand:2010} relaxed the conditional independence assumption by constructing a Bayesian model with a Gaussian dependence structure at the data level. However, while such an approach is computationally appealing, it lacks theoretical support for estimating joint tail probabilities---a key ingredient for assessing the risk of unprecedented spatial extreme events.} In the next sections, we introduce the natural extensions of the asymptotic GEV and GP distributions to the spatial context, namely max-stable processes and $r$-Pareto processes, respectively.

\subsection{Max-stable processes}\label{subsec:maxstable}
By analogy with \eqref{eq:maxstability}, a random process $Z(\bs{s})$, defined over a region $\bs{s}\in\calS\subset\Real^d$, is called max-stable if for any finite collection of sites $\bs{s}_1,\ldots,\bs{s}_D$, and any positive real $t>0$, there exist functions $\alpha_t(\bs{s})>0$ and $\beta_t(\bs{s})$ such that for all $z_1,\ldots,z_D$,
\begin{align}
\pr\{Z(\bs{s}_1)\leq \alpha_t(\bs{s}_1) z_1+ \beta_t(\bs{s}_1),\ldots, Z(\bs{s}_D)\leq & \alpha_t(\bs{s}_D) z_D+ \beta_t(\bs{s}_D)\}^t=\nonumber\\
&\pr\{Z(\bs{s}_1)\leq z_1,\ldots, Z(\bs{s}_D)\leq z_D\}.\label{eq:maxstabilityD}
\end{align}
By comparing \eqref{eq:maxstabilityD} to \eqref{eq:maxstability}, it is clear that max-stable processes have GEV margins. To focus on dependence, it is convenient to standardize the process to a common marginal scale. When the process $Z(\bs{s})$ has unit Fr\'echet margins, i.e., $\pr\{Z(\bs{s})\leq z\}=\exp(-1/z)$, $z>0$, which corresponds to the GEV distribution in \eqref{eq:GEV} with $\mu=\sigma=\xi=1$, then we have $\alpha_t(\bs{s}_j)=t$ and $\beta_t(\bs{s}_j)=0$, and \eqref{eq:maxstabilityD} may be simply written as
\begin{equation}\label{eq:maxstabilityDfrechet}
\pr\{Z(\bs{s}_1)\leq t z_1,\ldots, Z(\bs{s}_D)\leq t z_D\}^t=\pr\{Z(\bs{s}_1)\leq z_1,\ldots, Z(\bs{s}_D)\leq z_D\}.
\end{equation}
In the same way as the GEV distribution is the only possible {(non-degenerate)} limit for renormalized block maxima of independent and identically distributed random variables, max-stable processes are the only possible limit for renormalized pointwise maxima of random fields {with non-degenerate margins}. Specifically, let $Y_1(\bs{s}),Y_2(\bs{s}),\ldots$, $\bs{s}\in\calS\subset\Real^d$, be a sequence of independent and identically distributed random processes, and consider the process of pointwise maxima, $M_n(\bs{s})=\max_{i=1,\ldots,n} Y_i(\bs{s})$. If there exist sequences of functions $a_n(\bs{s})>0$ and $b_n(\bs{s})$ such that the renormalized process $a_n(\bs{s})^{-1}\{M_n(\bs{s})-b_n(\bs{s})\}$ converges as $n\to\infty$ to a process $Z(\bs{s})$ with non-degenerate margins, then the limit $Z(\bs{s})$ is a max-stable process satisfying \eqref{eq:maxstabilityD}. This asymptotic characterization has motivated the use of max-stable processes in practical extreme-value applications. 

\begin{figure}[t!]
\centering
 \includegraphics[width=\textwidth]{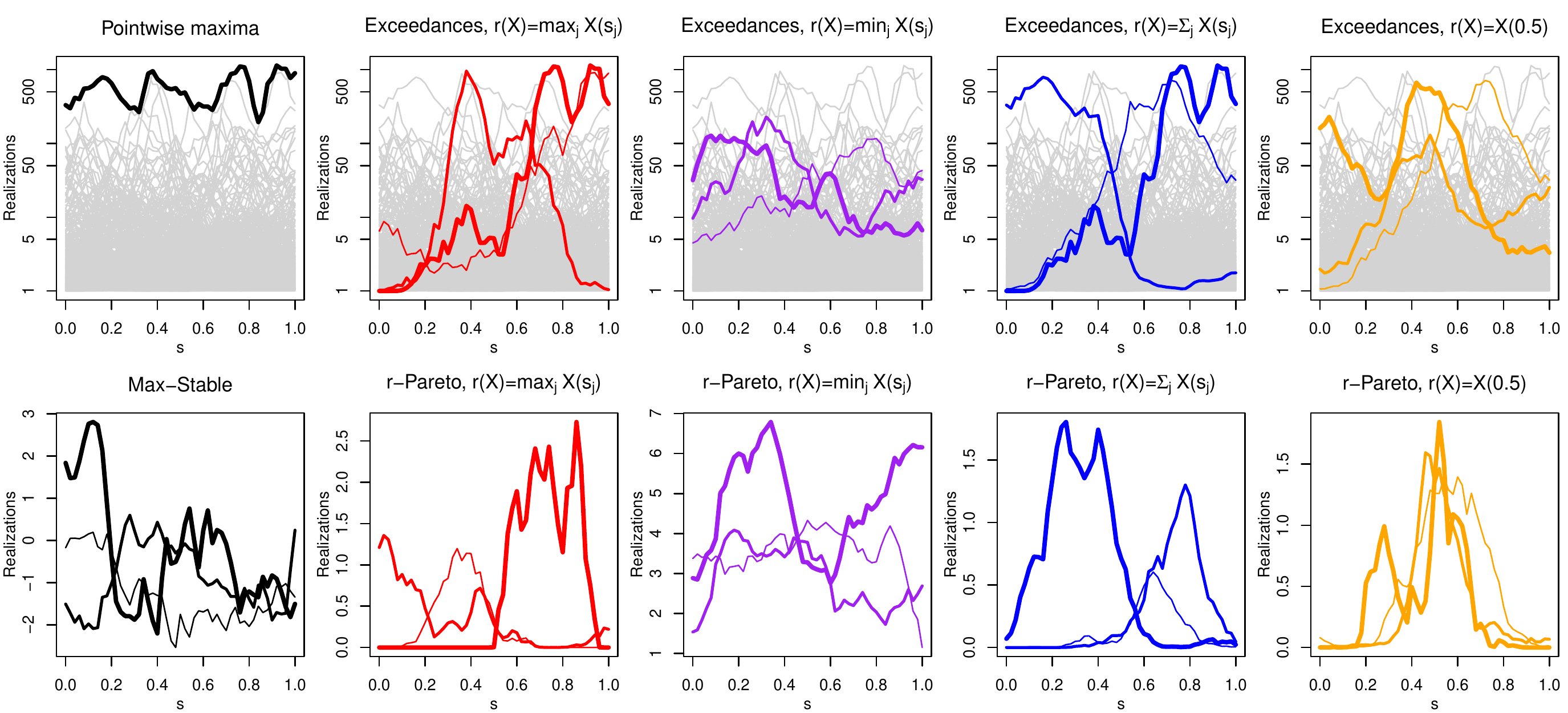}
  \caption{{Top row: independent realizations (grey) from the Gaussian scale mixture $X(s)=t\{RW(s)\}$, $s\in\calS=\{0,0.02,0.04,\ldots,1\}\subset[0,1]$, with $R\sim{\rm Pareto}(5)$ independent of the standard Gaussian process $W(s)$ with correlation function $\rho(s_1,s_2)=\exp\{-(|s_1-s_2|/0.4)^{1.5}\}$, and marginally transformed through $t(\cdot)$ such that $X(s)$ is on the unit Pareto scale. Highlighted curves are pointwise maxima ($1^\text{st}$ column, black) and the three largest $r$-exceedances based on the risk functionals $r(X)=\max_{s\in\calS}X(s)$ ($2^\text{nd}$ column, red), $r(X)=\min_{s\in\calS}X(s)$ ($3^\text{rd}$ column, purple), $r(X)=\sum_{s\in\calS}X(s){\rm d}s$ ($4^\text{th}$ column, blue), and $r(X)=X(0.5)$ ($5^\text{th}$ column, orange). The same random seed was used, so some $r$-exceedances may be identical in different panels. Bottom row: three independent realizations from the corresponding limiting (extremal-$t$) max-stable process of the form~\eqref{eq:maxstableprocess} ($1^\text{st}$ column) and $r$-Pareto processes of the form~\eqref{eq:rParetoprocess} ($2^\text{nd}$ to $5^\text{th}$ columns). For better visualization, processes in the top panels are displayed on a log-scale, while in the bottom panels, max-stable processes are plotted on the standard Gumbel scale, and $r$-Pareto processes have been transformed through the function $t(x)=\log\{1+x(e-1)\}$ such that $t(0)=0$ and $t(1)=1$. For $r$-exceedances and $r$-Pareto processes, thicker curves mean larger $r(X)$.}}
 \label{fig:rPareto}
\end{figure}

A useful way to build and characterize max-stable processes is via \citet{deHaan:1984}'s spectral representation; see also \citet{Schlather:2002} and \citet{deHaan.Ferreira:2006}, Chapter~9. Precisely, let $R_1,R_2,\ldots$ be points of a Poisson point process on $[0,+\infty]$ with intensity $r^{-2}{\rm d}r$, and $W_1(\bs{s}),W_2(\bs{s}),\ldots$ be independent copies of a non-negative stochastic process $W(\bs{s})\geq0$ with mean one, then max-stable processes with unit Fr\'echet margins may be constructed as follows:
\begin{equation}\label{eq:maxstableprocess}
Z(\bs{s})=\sup_{i=1,2\ldots} R_i W_i(\bs{s}).
\end{equation}
Essentially, max-stable processes can be seen as pointwise maxima of an infinite number of independent scale mixtures $R_iW_i(\bs{s})$, which may be interpreted as ``storms'' with overall amplitudes $R_i$ and spatial profiles $W_i(\bs{s})$. The heavy-tailedness of the power-law intensity of $\{R_i\}$ induces asymptotic dependence, and by construction max-stability. From \eqref{eq:maxstableprocess}, the finite-dimensional distributions $G$ of vectors $\{Z(\bs{s}_1),\ldots,Z(\bs{s}_D)\}^T$ have the form
\begin{equation}\label{eq:distmaxstable}
G(z_1,\ldots,z_D)=\pr\{Z(\bs{s}_1)\leq z_1,\ldots,Z(\bs{s}_D)\leq z_D\}=\exp\{-V(z_1,\ldots,z_D)\},
\end{equation}
where $V$ is known as the exponent function and may be expressed as $V(z_1,\ldots,z_D)=\E[\max\{W(\bs{s}_1)/z_1,\ldots,W(\bs{s}_D)/z_D\}]$. By specifying the $W$ process in different ways, various max-stable models can be constructed, the most popular of which include the Schlather model {and its refinements} \citep{Schlather:2002,Davison.Gholamrezaee:2012}, the Brown--Resnick model \citep{Brown.Resnick:1977,Kabluchko.etal:2009} and the extremal-$t$ model \citep{Opitz:2013a}. Another max-stable model with a different construction is the \citet{Reich.Shaby:2012a} model, which has gained popularity thanks to its conditional independence representation and its suitability for Bayesian inference in high dimensions. See \citet{Davison.etal:2012} and \citet{Davison.etal:2019} for a more detailed discussion of the pros and cons of such max-stable models, and a comparative study in concrete applications. Realizations from the extremal-$t$ process on a modified marginal scale are displayed in the bottom left panel of Figure~\ref{fig:rPareto}.

The main difficulty for using max-stable processes in practice is the complicated form of their likelihood function. From \eqref{eq:distmaxstable}, it can be deduced that the joint density function is
\begin{equation}\label{eq:densitymaxstable}
g(z_1,\ldots,z_D)=\exp\{-V(z_1,\ldots,z_D)\}\sum_{\pi\in\calP}\prod_{k=1}^{K}\left\{-V_{\pi_k}(z_1,\ldots,z_D)\right\},
\end{equation}
where $\calP$ denotes the collection of all partitions $\pi=\{\pi_1,\ldots,\pi_K\}$ of $\calD=\{1,\ldots,D\}$ with $\pi_k\subset\calD$, $k=1,\ldots,K\leq D$, and $V_{\pi_k}$ denotes the partial derivative of the exponent function $V$ with respect to the variables indexed by the set $\pi_k$; see, e.g., \citet{Castruccio.etal:2016}. The (full) likelihood function for independent replicates simply corresponds to a product of terms of the form \eqref{eq:densitymaxstable}. When $D$ is moderately large (i.e., roughly $D>5$--$10$), the general formula \eqref{eq:densitymaxstable} has too many terms to be used for likelihood inference in practice \citep{Padoan.etal:2010}. The use of event times, either explicitly or implicitly by integrating them out, may lead to some computational speed-up \citep{Stephenson.Tawn:2005,Davison.Gholamrezaee:2012,Wadsworth.Tawn:2014,Dombry.etal:2017,Huser.etal:2019}. However, while the ``explicit'' use of event times yields estimators that can be substantially biased \citep{Wadsworth:2015,Huser.etal:2016}, the ``implicit'' alternative approach is still computationally demanding in relatively low dimensions ($D\approx15$--$20$) with popular max-stable models \citep{Huser.etal:2019}. This ``full likelihood inference problem'' led \citet{Padoan.etal:2010} to propose using pairwise likelihoods instead, whereby pairwise densities of the form $\exp\{-V(z_i,z_j)\}\{V_1(z_i,z_j)V_2(z_i,z_j)-V_{12}(z_i,z_j)\}$ for all pairs of sites $\{\bs{s}_i,\bs{s}_j\}$ are combined together---potentially weighted---in an objective function by wrongly pretending that the pairs of variables $\{Z(\bs{s}_i),Z(\bs{s}_j)\}$ are mutually independent. Whilst this pairwise likelihood approach leads to valid inference (i.e., strong consistency, asymptotic normality) under mild regularity conditions \citep{Varin.Vidoni:2005,Padoan.etal:2010,Varin.etal:2011}, it entails a loss of information, which makes it less efficient than the maximum (full) likelihood approach, and it is also trickier to assess the uncertainty and to adapt it to the Bayesian framework. By contrast, approaches based on high threshold exceedances typically lead to simpler likelihood functions, as detailed in the next section.

\subsection{$r$-Pareto processes}\label{subsec:Pareto}

Max-stable processes form a natural analogue of the univariate GEV distribution through the operation of taking pointwise maxima. The natural analogue of the univariate GP distribution is less evident, because there is no unique way to extend the conditioning event $\{Y>u\}$ to the case of a spatial process $Y(\bs{s})$. Initial work in this line focused on conditioning events of the form $\{\sup_{\bs{s} \in \mathcal{S}}Y(\bs{s})>u\}$, in analogy with multivariate GP distributions \citep{Rootzen.Tajvidi:2006,Rootzen.etal:2018,Rootzen.etal:2018b}, and the resulting processes have been termed generalized Pareto processes. GP processes are a relatively recent addition to the spatial extreme-value literature. \citet{Buishand.etal:2008} provided a stochastic representation and properties, whilst \citet{Ferreira.deHaan:2014} gave a much more detailed study. Specifically, they considered the limiting distribution of 
\begin{align}
\left[1+\xi(\bs{s})\left\{{Y(\bs{s})-b_n(\bs{s})\over a_n(\bs{s})}\right\}\right]^{1/\xi(\bs{s})}_+ \mid \sup_{\bs{s} \in \mathcal{S}} {Y(\bs{s})-b_n(\bs{s})\over a_n(\bs{s})}>0, \label{eq:gpprocconv}
\end{align}
where $a_n(\bs{s})>0, b_{n}(\bs{s})$ are as in the convergence of max-stable processes, $\xi(\bs{s})$ is the shape parameter of the GEV or GP distribution at site $\bs{s}$, and $a_+ = \max(a,0)$. Theory and practice appear simpler when marginal distributions are pre-transformed, and a common choice is standard Pareto: for any $\bs{s}_j \in \mathcal{S}$, $\Pr\{\tilde{Y}(\bs{s}_j)>y\} = y^{-1}$, $y\geq1$. Here, $\tilde{Y}(\bs{s})$ denotes a standardized version of the process $Y(\bs{s})$. In this case, \eqref{eq:gpprocconv} becomes ${\tilde{Y}(\bs{s})/n \mid \sup_{\bs{s} \in \mathcal{S}}\tilde{Y}(\bs{s})>n}$, and the limit as $n \to \infty$ is called a standard Pareto process. Such a transformation to $\tilde{Y}(\bs{s})$ also leads to max-stable processes with unit Fr\'echet margins. 

Notice that while max-stable processes have finite-dimensional margins that are multivariate max-stable distributions, the same is not true of generalized Pareto processes, since the conditioning event relates to the full infinite-dimensional process. This is the same reason that lower $D'<D$ dimensional margins of $D$-dimensional multivariate generalized Pareto distributions are not, in general, multivariate generalized Pareto distributed. However, the marginal distributions conditional upon an exceedance within the marginal index set, are multivariate generalized Pareto.

The conditioning event $\{\sup_{\bs{s} \in \mathcal{S}}\tilde{Y}(\bs{s})>n\}$ has the drawback that theoretically one needs to observe $\tilde{Y}(\bs{s})$ over its entire domain $\mathcal{S}$. \citet{Dombry.Ribatet:2015} introduced an extension to GP processes, which they termed $\ell$-Pareto processes, but which have also been called $r$-Pareto processes by \citet{deFondeville.Davison:2018}. The quantity $\ell$ or $r$ has been variously termed the cost or risk functional, and determines the specific definition of a spatial extreme event. It takes the whole process $\tilde{Y}(\bs{s})$ as input and returns a positive scalar, which corresponds to the ``size'' or ``severity'' of the process $\tilde{Y}(\bs{s})$. We adopt the $r$ notation here. The functional $r$ should be homogeneous of order 1, i.e., for any $c>0$, $r(cY)=cr(Y)$. This offers potential to condition on other events of interest, such as $r(\tilde{Y}) = \int_\calS\tilde{Y}(\bs{s}) \mathrm{d}\bs{s}$, $r(\tilde{Y})=\inf_{\bs{s}\in\calS}\tilde{Y}(\bs{s})$, or quantities involving finite observation domains: $r(\tilde{Y}) = \max_{1\leq j \leq D} \tilde{Y}(\bs{s}_j)$, $r(\tilde{Y}) = \sum_{1\leq j \leq D} \tilde{Y}(\bs{s}_j)$, $r(\tilde{Y}) = \min_{1\leq j \leq D} \tilde{Y}(\bs{s}_j)$ or even $r(\tilde{Y}) = \tilde{Y}(\bs{s}_0)$ for some site $\bs{s}_0$.  The choice $r(\tilde{Y}) = \sup_{\bs{s} \in \calS}\tilde{Y}(\bs{s})$ is also valid, leading to standard Pareto processes as related to GP processes mentioned above. Various types of risk functionals $r$, and some corresponding $r$-exceedances are illustrated in Figure~\ref{fig:rPareto}.

{For a homogeneous risk functional $r$, an $r$-Pareto process can be expressed as 
\begin{align}\label{eq:rParetoprocess}
X(\bs{s}) = R\,W(\bs{s}),
\end{align}
 $R \sim \text{Pareto}(1)$ is independent of $W(\bs{s}) \geq 0$, which satisfies $r(W) = 1$ almost surely. The process~\eqref{eq:rParetoprocess} has a threshold-stability property analogous to the max-stability property in~\eqref{eq:maxstabilityDfrechet}; specifically for suitable sets $B$, and thresholds $v \geq 1$,
 \begin{align}
 \Pr\{X(\bs{s})/v \in B \mid r(X)>v\} = \Pr\{X(\bs{s}) \in B\}. \label{eq:threshstab}
 \end{align}
 
If we have a process $\tilde{Y}(\bs{s})$ with standard Pareto margins, then we may suppose that for large enough thresholds $u$ on the Pareto scale, ${\tilde{Y}(\bs{s})/u \mid r(\tilde{Y}) >u} \overset{D}{\approx} X(\bs{s})$. Selection of an appropriate distribution for $W(\bs{s})$ will lead to a model for $X(\bs{s})$ and hence for $r$-exceedances of $\tilde{Y}(\bs{s})$, i.e., spatial events such that $r(\tilde{Y})>u$. The construction~\eqref{eq:rParetoprocess} is used in Figure~\ref{fig:rPareto} to generate realizations from $r$-Pareto processes.} 
 
{To obtain a process $W(\bs{s})$ in~\eqref{eq:rParetoprocess} such that $r(W)=1$, one possibility is to begin with another process $Q(\bs{s})\geq 0$ for which $r(Q)>0$ and set $W(\bs{s})=Q(\bs{s})/r(Q)$. Typically, one might be interested in $r$-Pareto processes that correspond to known max-stable models and computable likelihoods, which requires specific choices.} Similarly to univariate and multivariate extremes, $r$-Pareto and max-stable processes can be linked via a Poisson process representation \citep{Thibaud.Opitz:2015}, and likelihood-based inference and simulation algorithms for $r$-Pareto equivalents of the extremal-$t$ and Brown--Resnick process are detailed respectively in \citet{Thibaud.Opitz:2015} and \citet{deFondeville.Davison:2018}. To obtain likelihoods, calculation of the relevant density of the Poisson mean measure is required, along with normalization constants that are determined by the mean measure and the form of $r$. The former is equivalent to calculation of partial derivatives of the exponent function $V$ for a max-stable process, where attention needs to be paid to discontinuities in the measure for extremal-$t$ processes \citep{Thibaud.Opitz:2015}. This issue can be circumvented by use of a censored likelihood, which also protects against bias in inference that can be caused by non-extreme values \citep{Huser.etal:2016}, and has also been advocated by \citet{Wadsworth.Tawn:2014}. As elsewhere, the process is assumed to be observed at $D$ spatial locations. The general form for a censored Pareto process likelihood for $n$ independent replicates with $r(\tilde{Y})>u$ is
\begin{align}\label{eq:likelihoodPareto}
 \prod_{i=1}^{n}\left[ -{V_{I_i}\{\max(\tilde{\bs{y}}_i,u)\} \over K_{r}(u)}\right],
\end{align}
where $I_i =\{j: \tilde{Y}_i(\bs{s}_j)>u\}\subseteq\{1,\ldots,D\}$, $V_{I}$ is the partial derivative of $V$ with respect to all components in $I$, and $\max(\tilde{\bs{y}}_i,u)$ is the $D$-dimensional vector consisting of elements $\tilde{y}_{i}(\bs{s}_j)$ where $\tilde{y}_{i}(\bs{s}_j)>u$ and $u$ where $\tilde{y}_{i}(\bs{s}_j)<u$. The quantity $K_{r}(u)$ is the normalization constant. When $r(\tilde{Y}) = \max_{1\leq j \leq D} \tilde{Y}(\bs{s}_j)$, then $K_r(u) = V(u,\ldots,u)$; if $r(\tilde{Y}) = \sum_{j=1}^D \tilde{Y}(\bs{s}_j)/D$, then $K_r(u)$ does not depend on any model parameters, which simplifies the inference. By comparing \eqref{eq:likelihoodPareto} with \eqref{eq:densitymaxstable}, we notice that the number of terms and partial derivatives of $V$ to compute is much smaller with \eqref{eq:likelihoodPareto}, which makes $r$-Pareto processes amenable to vastly higher-dimensional inference than max-stable processes.

Nevertheless, the likelihood function \eqref{eq:likelihoodPareto} may still be burdensome to compute for two main reasons: first, the expression of $V$ and its partial derivatives for the popular Brown--Resnick and extremal-$t$ models involve calculation of multivariate Gaussian or $t$ distribution functions up to dimension $D-1$, which has a prohibitive effect on the number of observation locations that can be used. Second, the normalizing constant $K_r(u)$ in \eqref{eq:likelihoodPareto} may be awkward to compute for general risk functionals $r$. To circumvent these issues, \citet{deFondeville.Davison:2018} proposed two possible remedies: (i) more rapid calculation of multivariate Gaussian or $t$ distribution functions via Quasi-Monte Carlo techniques; and (ii) use of a gradient score algorithm in place of maximum likelihood. The latter avoids the calculation of normalizing constants, while the use of weighting functions can also circumvent the need for censoring and hugely reduce the computational burden. \citet{deFondeville.Davison:2018} applied this methodology, implemented in the R package \texttt{mvPot} \citep{deFondeville.Belzile:2018}, to a dataset of satellite rainfall measurements with $D=3600$.

Although $r$-Pareto processes appear to generalize GP processes to alternative definitions of spatial extremes in a natural way, there are nonetheless serious practical drawbacks if the event of interest cannot naturally be expressed on the standardized scale of $\tilde{Y}(\bs{s})$. For example, if $Y(\bs{s})$ represents rainfall and one wishes to condition on a large value of the areal rainfall, then the conditioning quantity of interest is $\int_{\mathcal{S}} Y(\bs{s}) \mathrm{d}\bs{s}$, not $\int_{\mathcal{S}} \tilde{Y}(\bs{s}) \mathrm{d}\bs{s}$. 
{Building on \citet{deFondeville.Davison:2018}, current research focuses on developing} generalized $r$-Pareto processes to allow consideration of events on their original scale, subject to the condition that the shape parameter $\xi(\bs{s})$ is constant over space.

\section{Sub-asymptotic models for spatial extremes}\label{sec:AIAD}

\subsection{Asymptotic dependence classes}\label{subsec:AIAD}
To characterize the strength of extremal dependence in a process $Y(\bs{s})$, $\bs{s}\in\calS$, we can consider the bivariate $\chi$-measure, defined in \eqref{eq:chi} as
\begin{equation}\label{eq:chi2}
\chi(\bs{s}_1,\bs{s}_2)=\lim_{u\to1}\chi_u(\bs{s}_1,\bs{s}_2)=\lim_{u\to1}\pr\{Y(\bs{s}_1)>F_1^{-1}(u)\mid Y(\bs{s}_2)>F_2^{-1}(u)\},
\end{equation}
where $Y(\bs{s}_1)\sim F_1$ and $Y(\bs{s}_2)\sim F_2$, such that $F_1\{Y(\bs{s}_1)\},F_2\{Y(\bs{s}_2)\}\sim{\rm Unif}(0,1)$ when $Y$ has continuous margins, and $u\in(0,1)$ in \eqref{eq:chi2} is a quantile on the uniform scale. In the copula literature, the $\chi$-measure is often denoted by the symbol $\lambda$ and called the \emph{coefficient of upper tail dependence}. Asymptotic dependence arises when $\chi(\bs{s}_1,\bs{s}_2)>0$, whereas asymptotic independence corresponds to $\chi(\bs{s}_1,\bs{s}_2)=0$. For Pareto processes, we have $\chi_u(\bs{s}_1,\bs{s}_2)={2-V(1,1)}$ for all $u$ above a certain level, with $V$ the bivariate exponent function corresponding to the pair of sites $\{\bs{s}_1,\bs{s}_2\}$, while for max-stable processes,
	\begin{align}
\chi_u(\bs{s}_1,\bs{s}_2)=2-V(1,1)+\calO(1-u), \qquad u\to1. \label{eq:MSchiu}
	 \end{align}
 This implies that these asymptotic extreme-value processes cannot adequately reflect situations where the dependence strength weakens as events become more extreme, and that they are always asymptotically dependent, unless exactly independent. This is a major limitation in practice, especially in environmental applications, where it is often found that the most severe spatial extreme events are more localized {\citep{Huser.Wadsworth:2019}}.

Because of the importance of the asymptotic independence case in practice, it is useful to additionally consider the rate at which the sub-asymptotic $\chi$-measure, $\chi_u(\bs{s}_1,\bs{s}_2)$, in \eqref{eq:chi} and \eqref{eq:chi2} tends to zero as $u\to1$. Following \citet{Ledford.Tawn:1996}, we may assume that 
\begin{equation}\label{eq:eta}
\chi_u(\bs{s}_1,\bs{s}_2)\sim \mathcal{L}\{(1-u)^{-1}\} (1-u)^{1/\eta(\bs{s}_1,\bs{s}_2)-1},\qquad u\to1,
\end{equation}
where $\mathcal{L}$ is a slowly-varying function at infinity, i.e., $\mathcal{L}(tx)/\mathcal{L}(x)\to1$ as $x\to\infty$ for all $t>0$, and $\eta(\bs{s}_1,\bs{s}_2)\in(0,1]$ is called the \emph{coefficient of tail dependence}, also known as the \emph{coefficient of residual tail dependence}. Notice that in the copula literature, \citet{Hua.Joe:2011} similarly defined the quantity $\kappa(\bs{s}_1,\bs{s}_2)=1/\eta(\bs{s}_1,\bs{s}_2)$ as the \emph{tail order}. While the value of $\chi(\bs{s}_1,\bs{s}_2)$ characterizes the asymptotic dependence class, $\eta(\bs{s}_1,\bs{s}_2)$ determines the flexibility of a model to capture the sub-asymptotic joint tail behavior and is principally used in the asymptotic independence case. Specifically, when $\eta(\bs{s}_1,\bs{s}_2)<1$, we get asymptotic independence, which may be further classified into (i) positive association with $\eta(\bs{s}_1,\bs{s}_2)>1/2$; (ii) near-independence with $\eta(\bs{s}_1,\bs{s}_2)=1/2$; and (iii) negative association with $\eta(\bs{s}_1,\bs{s}_2)<1/2$. The only case corresponding to asymptotic dependence is when $\eta(\bs{s}_1,\bs{s}_2)=1$ and $\mathcal{L}(x)$ has a positive limit as $x\to\infty$. The case $\eta(\bs{s}_1,\bs{s}_2)=1$ with $\mathcal{L}(x)\to0$ as $x\to\infty$ is a subtle boundary case leading to asymptotic independence, which is rarely encountered in practice (but see \citet{Huser.Wadsworth:2019} for an example).

\begin{figure}[t!]
\centering
 \includegraphics[width=0.8\textwidth]{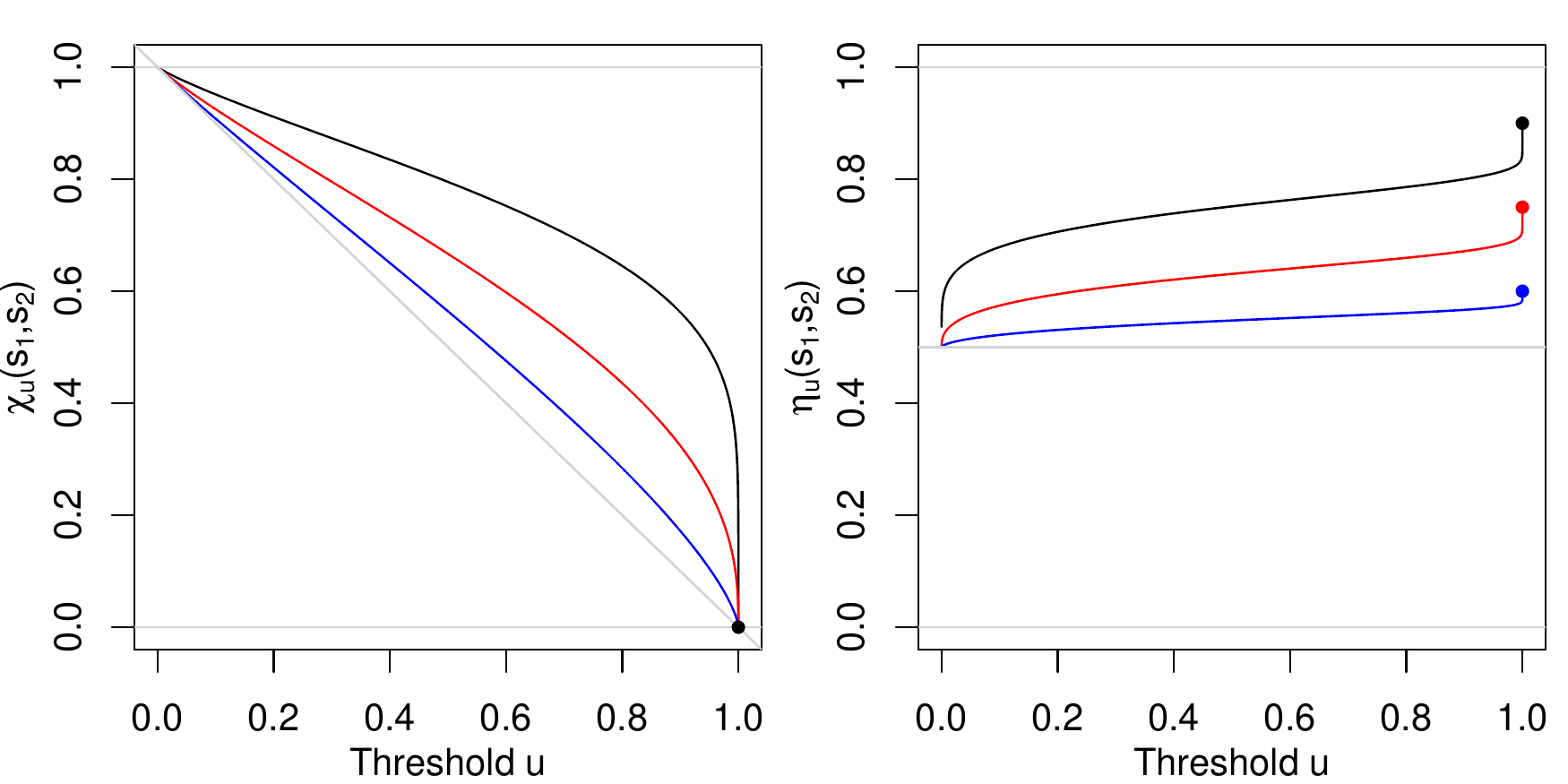}
  \caption{Sub-asymptotic $\chi$ (left) and $\eta$ (right) measures, namely $\chi_u(\bs{s}_1,\bs{s}_2)$ in \eqref{eq:chi2} and $\eta_u(\bs{s}_1,\bs{s}_2)=\log(1-u)/\log\pr\{Y(\bs{s}_1)>F_1^{-1}(u), Y(\bs{s}_2)>F_2^{-1}(u)\}$, respectively, plotted with respect to quantile level $u\in(0,1)$, as well as the limiting quantities $\chi(\bs{s}_1,\bs{s}_2)=\lim_{u\to1}\chi_u(\bs{s}_1,\bs{s}_2)$ and $\eta(\bs{s}_1,\bs{s}_2)=\lim_{u\to1}\eta_u(\bs{s}_1,\bs{s}_2)$ (small dots), for a Gaussian process $Y(\bs{s})$ and correlation $0.2$ (blue), $0.5$ (red) and $0.8$ (black) for the random vector $\{Y(\bs{s}_1),Y(\bs{s}_2)\}^T$.}
 \label{fig:chisGaussian}
\end{figure}

The most popular and widely-used class of asymptotic independence models are Gaussian processes and their marginally transformed counterparts (i.e., so-called ``trans-Gaussian'' processes). {Such Gaussian-based models have been widely used for spatial modeling; see, e.g., \citet{Sang.Gelfand:2010} for an environmental application.} (Trans-)Gaussian processes with underlying correlation function $\rho(\bs{s}_1,\bs{s}_2)$ satisfy \eqref{eq:eta} with $\eta(\bs{s}_1,\bs{s}_2)=\{1+\rho(\bs{s}_1,\bs{s}_2)\}/2$ \citep{Sibuya:1960,Ledford.Tawn:1996}, so that $\eta(\bs{s}_1,\bs{s}_2)<1$ (asymptotic independence) whenever $\rho(\bs{s}_1,\bs{s}_2)<1$. The only asymptotic dependence scenario is when $\rho(\bs{s}_1,\bs{s}_2)=1$ (perfect dependence). This result shows that with Gaussian processes, the correlation function $\rho(\bs{s}_1,\bs{s}_2)$ controls the decay of dependence both with respect to spatial distance, and with respect to quantile level. Thus, for fixed correlation $\rho(\bs{s}_1,\bs{s}_2)$, the joint tail decay rate is fixed. Gaussian processes are therefore rather rigid for modeling asymptotically independent extremes, and there are not many flexible alternatives. To illustrate these concepts and the rigidity of the Gaussian dependence structure, Figure~\ref{fig:chisGaussian} shows the sub-asymptotic $\chi$- and $\eta$-measures, namely $\chi_u(\bs{s}_1,\bs{s}_2)$ in \eqref{eq:chi2} and $\eta_u(\bs{s}_1,\bs{s}_2)=\log(1-u)/\log\pr\{Y(\bs{s}_1)>F_1^{-1}(u), Y(\bs{s}_2)>F_2^{-1}(u)\}$, respectively, as well as the limiting quantities $\chi(\bs{s}_1,\bs{s}_2)=\lim_{u\to1}\chi_u(\bs{s}_1,\bs{s}_2)$ and $\eta(\bs{s}_1,\bs{s}_2)=\lim_{u\to1}\eta_u(\bs{s}_1,\bs{s}_2)$, for a Gaussian process $Y(\bs{s})$ with various correlations. While $\chi_u(\bs{s}_1,\bs{s}_2)$ converges to zero as $u\to1$ whatever the correlation, $\eta_u(\bs{s}_1,\bs{s}_2)$ tends to a constant less than one, which implies asymptotic independence. Notice that the value of $\eta(\bs{s}_1,\bs{s}_2)$ is larger for higher correlation, indicating a slower convergence rate to the limit.

In the following subsections, we describe various types of spatial models that can more flexibly capture the ``sub-asymptotic'' extremal behavior. In Section~\ref{subsec:invertedMS}, we focus on inverted max-stable models, which are asymptotically independent, and their hybrid max-mixture extensions. In Section~\ref{subsec:RandomScale}, we describe random scale mixtures and related models that are specifically designed for bridging the asymptotic dependence and independence regimes. In Section~\ref{subsec:MaxID}, we describe recently proposed max-infinitely divisible models that can capture asymptotic independence in block maxima data, while keeping a popular (asymptotically dependent) max-stable model on the boundary of the parameter space.

\subsection{Inverted max-stable processes and max-mixture models}\label{subsec:invertedMS}
\citet{Wadsworth.Tawn:2012b} proposed an alternative class of asymptotic independence models for spatial extremes, that are generally more flexible than Gaussian processes, at the price of being more tricky to fit. Specifically, they introduced the wide class of inverted max-stable processes, constructed by ``swapping'' the tails of a max-stable process. More precisely, let $Z(\bs{s})$, $\bs{s}\in\calS$, be a max-stable process with unit Fr\'echet margins defined as in \eqref{eq:maxstableprocess}, and characterized by the exponent function $V$. Then, the corresponding inverted max-stable (IMS) process is simply defined as $Z^{\rm IMS}(\bs{s})=1/Z(\bs{s})$, which has therefore unit exponential margins, i.e., $\pr\{Z^{\rm IMS}(\bs{s})>z\}=\exp(-z)$, $z>0$. By noticing that the bivariate survival function is
\begin{equation}\label{eq:invertedMSjoint}
\pr\{Z^{\rm IMS}(\bs{s}_1)>z_1,Z^{\rm IMS}(\bs{s}_2)>z_2\}=\exp\{-V(1/z_1,1/z_2)\},
\end{equation}
where $V$ here denotes the bivariate restriction of the exponent function to the pair of variables $\{Z(\bs{s}_1),Z(\bs{s}_2)\}$, it can be shown that the process $Z^{\rm IMS}(\bs{s})$ satisfies \eqref{eq:eta} with $\eta(\bs{s}_1,\bs{s}_2)=1/V(1,1)$. The quantity $\theta(\bs{s}_1,\bs{s}_2)=V(1,1)\in[1,2]$ is known as the bivariate extremal coefficient of the max-stable vector $\{Z(\bs{s}_1),Z(\bs{s}_2)\}^T$; see, e.g., \citet{Schlather.Tawn:2003}, \citet{Davison.etal:2012} and \citet{Davison.etal:2019}. Hence, for each (asymptotically dependent) max-stable process, there exists an asymptotically independent inverted max-stable counterpart with $\eta(\bs{s}_1,\bs{s}_2)\in[1/2,1]$. {This implies in particular that inverted max-stable processes are positively associated.} As max-stable dependence structures are highly non-Gaussian and may potentially be asymmetric, such a construction substantially widens the class of possible asymptotic independence models. Likelihood inference for inverted max-stable processes suffers the same limitations as max-stable processes themselves. Thus, by analogy with the max-stable framework (recall Section~\ref{subsec:maxstable} and the joint density \eqref{eq:densitymaxstable}), the pairwise likelihood approach is commonly used by combining pairwise likelihood contributions of the form $\exp\{-V(1/z_1,1/z_2)\}\{V_1(1/z_1,1/z_2)V_2(1/z_1,1/z_2)-V_{12}(1/z_1,1/z_2)\}(z_1z_2)^{-2}$, obtained by differentiating \eqref{eq:invertedMSjoint} with respect to $z_1$ and $z_2$.

In the same way as (non-trivial) max-stable processes are always asymptotically dependent, inverted max-stable processes are always asymptotically independent. \citet{Wadsworth.Tawn:2012b} were the first to propose hybrid models that combine these two asymptotic regimes in a unified framework. Let $Z_1(\bs{s})$ and $\tilde{Z}_2(\bs{s})$, $\bs{s}\in\calS$, be two independent max-stable processes with unit Fr\'echet margins, and define $Z_2(\bs{s})=-1/\log[1-\exp\{-1/\tilde{Z}_2(\bs{s})\}]$ as the inverted max-stable counterpart of $\tilde{Z}_2(\bs{s})$ but on the unit Fr\'echet scale. Then, a max-mixture model is defined as the pointwise maximum
\begin{equation}\label{eq:maxmixture}
Z(\bs{s})=\max\{aZ_1(\bs{s}),(1-a)Z_2(\bs{s})\},
\end{equation}
where the parameter $a\in[0,1]$ controls the ``mixture proportion'' between the max-stable versus inverted max-stable processes in \eqref{eq:maxmixture}. When $a=1$, the resulting max-mixture process $Z(\bs{s})$ reduces to $Z_1(\bs{s})$ (max-stable) and when $a=0$, it reduces to $Z_2(\bs{s})$ (inverted max-stable). Moreover, if the max-stable process $Z_1(\bs{s})$ has dependence only up to a finite spatial distance $h^\star<\infty$ (and is independent beyond $h^\star$), then for any $a\in(0,1)$ the max-mixture process $Z(\bs{s})$ has the appealing and intuitive property of being asymptotically dependent for short distances $h=\|\bs{s}_1-\bs{s}_2\|<h^\star$ and asymptotically independent (but not necessarily exactly independent) for $h\geq h^\star$. Moreover, even when $h^\star = \infty$, the inverted max-stable component gives flexibility in the rate $\chi_u(\bs{s}_1,\bs{s}_2)-\chi(\bs{s}_1,\bs{s}_2)$ as $u\to\infty$ so that it may differ from~\eqref{eq:MSchiu} and improves model fit at sub-asymptotic levels. The same construction may be used by replacing $Z_1$ by any asymptotically dependent process and $Z_2$ by any asymptotically independent process. The bivariate distribution stemming from \eqref{eq:maxmixture} can be conveniently expressed as the product $\pr\{Z(\bs{s}_1)\leq z_1,Z(\bs{s}_2)\leq z_2\}=\pr\{Z_1(\bs{s}_1)\leq z_1/a,Z_1(\bs{s}_2)\leq z_2/a\}\pr\{Z_2(\bs{s}_1)\leq z_1/(1-a),Z_2(\bs{s}_2)\leq z_2/(1-a)\}$, which may be exploited for pairwise likelihood inference, usually based on high threshold exceedances by censoring low values. This model has been used, e.g., by \citet{Bacro.etal:2016} and \citet{Ahmed.etal:2019}, 
but it has the drawback of being usually quite heavily parametrized and that estimation of the crucial parameter $a$ is difficult. In the next section, we present more parsimonious spatial extreme-value models that can also capture both asymptotic dependence regimes.

\subsection{Random scale mixtures, and related models}\label{subsec:RandomScale}
Both max-stable and Pareto processes are built from scale mixtures of the form $X(\bs{s})=RW(\bs{s})$, where the common, spatially-constant random factor $R$ has a heavy Pareto tail. In the case of max-stable processes constructed as in \eqref{eq:maxstableprocess}, the Poisson points $\{R_i\}$ indeed have a power-law intensity $r^{-2}{\rm d}r$ on $[0,+\infty]$, whilst in the case of Pareto processes in \eqref{eq:rParetoprocess}, the random variable $R$ has the unit Pareto distribution with density function $r^{-2}$ on $[1,+\infty]$. Intuitively, this heavy-tail behavior creates extreme ``shocks'' in the randomly scaled mixture $RW(\bs{s})$, which ``uplifts'' the whole process simultaneously and creates co-occurrences of extreme events at multiple locations. For comparatively light-tailed $W$---ensured by finite first moment---this mechanism yields asymptotic dependence.

In order to get more flexible families of extremal dependence structures, we can consider random scale mixture constructions $X(\bs{s})=RW(\bs{s})$ with general $R$ and/or $W$. The extremal dependence properties of such models have been almost completely characterized in the bivariate case by \citet{Engelke.etal:2019} (see also the references therein), and we now focus on three especially interesting cases that can bridge asymptotic dependence classes, {namely certain types of Gaussian scale mixtures \citep{Huser.etal:2017}, Gaussian location mixtures \citep{Krupskii.etal:2018}, and the \citet{Huser.Wadsworth:2019} model. These models are introduced in more detail below.}

The first interesting case is to consider (elliptically-contoured) Gaussian scale mixtures \citep{Huser.etal:2017}, where $R\geq0$ has some distribution $F_R$ on $[0,+\infty]$, and $W(\bs{s})$ is a standard Gaussian process with correlation function $\rho(\bs{s}_1,\bs{s}_2)$, independent of $R$. When $R=r_0>0$ almost surely, we obtain Gaussian processes with asymptotic independence, but more flexible models may be obtained by considering mixing variables $R$ with Pareto or Weibull-like tail decay. Specifically, assume that $1-F_R$ is regularly-varying at infinity, or equivalently that $F_R$ is Pareto-tailed, i.e.,
\begin{equation}\label{eq:ParetoTailed}
1-F_R(r)\sim K\, r^{-\gamma},\qquad r\to\infty,
\end{equation}
where $K>0$ is a positive constant and $\gamma>0$ determines the power-law tail decay rate. Then, we can show that the Gaussian scale mixture $X(\bs{s})$ is asymptotically dependent with $\chi(\bs{s}_1,\bs{s}_2)=2-2T_{\gamma+1}[(1+\gamma)^{1/2}\{1-\rho(\bs{s}_1,\bs{s}_2)\}\{1-\rho(\bs{s}_1,\bs{s}_2)^2\}^{-1/2}]$ and $\eta(\bs{s}_1,\bs{s}_2)=1$, where $T_\nu$ denotes the Student's $t$ distribution function with $\nu>0$ degrees of freedom \citep{Huser.etal:2017}. This case includes for example Student's $t$ processes, constructed by taking $R$ as a specific inverse-gamma random variable. While the correlation function $\rho(\bs{s}_1,\bs{s}_2)$ mostly controls the decay of dependence with spatial distance, the additional parameter $\gamma$ adds substantial flexibility to capture different levels of asymptotic dependence for fixed correlation. Alternatively, instead of \eqref{eq:ParetoTailed}, we may assume that $F_R$ is Weibull-tailed, i.e.,
\begin{equation}\label{eq:WeibullTailed}
1-F_R(r)\sim K\, r^{\alpha}\exp(-\theta r^\beta),\qquad r\to\infty,
\end{equation}
where $K>0$, $\alpha\in\Real$, $\theta>0$ and $\beta>0$. The Weibull index $\beta$ now determines the tail decay rate. In this case, we can show that the Gaussian scale mixture $X(\bs{s})$ is asymptotically independent with $\chi(\bs{s}_1,\bs{s}_2)=0$ and $\eta(\bs{s}_1,\bs{s}_2)=[\{1+\rho(\bs{s}_1,\bs{s}_2)\}/2]^{\beta/(\beta+2)}$ \citep{Huser.etal:2017}. The Gaussian case can be viewed as a special limiting case obtained as $\beta\to\infty$, while Laplace random fields have $\beta=2$ \citep{Opitz:2016}, but by treating $\beta>0$ as an additional free parameter, we considerably increase the flexibility to capture the sub-asymptotic behavior. When the extremal dependence class is unclear, \citet{Huser.etal:2017} proposed combining the Pareto-tailed and Weibull-tailed frameworks in \eqref{eq:ParetoTailed} and \eqref{eq:WeibullTailed}, respectively, using the model
\begin{equation}\label{eq:Huseretal}
F_R(r)= 1-\exp\{-\gamma (r^\beta-1)/\beta\},\qquad r\geq1,
\end{equation}
for $\gamma>0$ and $\beta>0$, which is Weibull-tailed when $\beta>0$ and converges to the Pareto distribution $1-r^{-\gamma}$, $r\geq1$, as $\beta\downarrow0$. This model is therefore asymptotically independent, but it keeps a flexible asymptotically dependent submodel on the boundary of the parameter space. Realizations for $\beta\downarrow0$ are displayed in the top panels of Figure~\ref{fig:rPareto}. By conditioning on $R$, it can be easily verified that the general form of the distribution function for random scale mixture models $X(\bs{s})=RW(\bs{s})$ may be expressed as
\begin{equation}\label{eq:RWCDF}
\pr\{X(\bs{s}_1)\leq x_1,\ldots,\leq,X(\bs{s}_D)\leq x_D\}=\int_0^\infty F_{\bs{W}}(x_1/r,\ldots,x_D/r) {\rm d}F_R(r),
\end{equation}
where $F_{\bs{W}}$ is the joint distribution of the vector $\bs{W}=\{W(\bs{s}_1),\ldots,W(\bs{s}_D)\}^T$, while the joint density function may be obtained by differentiating \eqref{eq:RWCDF} under the integral sign. \citet{Huser.etal:2017} showed how to exploit \eqref{eq:RWCDF} to perform censored likelihood inference based on high threshold exceedances for this class of models, but this remains fairly intensive in moderate dimensions (roughly $D>30$) in cases where the random variable $R$ cannot be integrated out in explicit form and (uni-dimensional) numerical integrals are thus required. 
Another appealing property of Gaussian scale mixture models is that they are easily amenable to unconditional or conditional simulation, which is typically required for the evaluation of spatial risk measures and for spatial prediction. 

A second related class of models is to consider Gaussian location mixtures \citep{Krupskii.etal:2018}, which can be viewed as a special type of random scale mixture after exponentiation---a monotone marginal transformation keeping the dependence structure intact. Such models are defined as $\tilde{X}(\bs{s})=\tilde{R} + \tilde{W}(\bs{s})$, where $\tilde{R}$ is a random variable with distribution $F_{\tilde{R}}$ on the whole real line and $\tilde{W}(\bs{s})$ is a standard Gaussian process with correlation function $\rho(\bs{s}_1,\bs{s}_2)$, independent of $\tilde{R}$. In this case, we can show that if $F_{\tilde{R}}$ is Pareto-tailed in the sense of \eqref{eq:ParetoTailed} or Weibull-tailed in the sense of \eqref{eq:WeibullTailed} with $\beta<1$, then we get perfect asymptotic dependence, i.e., $\chi(\bs{s}_1,\bs{s}_2)=1$. If, however, $F_{\tilde{R}}$ is exponential-tailed (i.e., Weibull-tailed as in \eqref{eq:WeibullTailed} with $\beta=1$), then we get asymptotic dependence with $\chi(\bs{s}_1,\bs{s}_2)=2-2\Phi(\theta[\{1-\rho(\bs{s}_1,\bs{s}_2)\}/2]^{1/2})$, where $\Phi$ denotes the standard normal distribution function. Finally, when $F_{\tilde{R}}$ is Weibull-tailed with $\beta>1$, we get asymptotic independence with $\chi(\bs{s}_1,\bs{s}_2)=0$. Further theoretical results, as well as modeling and inference considerations are detailed in \citet{Krupskii.etal:2018}. See also \citet{Krupskii.Joe:2013} for general theory on this kind of models in the multivariate case, \citet{Krupskii.Genton:2017} for the extension to the spatio-temporal framework, and \citet{CastroCamilo.Huser:2020} for an application in the non-stationary spatial context.

In Gaussian scale or location mixtures discussed above, the random variable $R$ (or $\tilde{R}$) and the process $W(\bs{s})$ (or $\tilde{W}(\bs{s})$) are defined on fundamentally different marginal scales, which makes it impossible to bridge asymptotic dependence and independence in the interior of the parameter space. To achieve this, and to allow easy inference on the asymptotic dependence class, \citet{Huser.Wadsworth:2019} {proposed an alternative random scale mixture model} defined as $R^{\delta}W(\bs{s})^{1-\delta}$ for some parameter $\delta\in[0,1]$, where $R\geq1$ is a unit Pareto random variable, and $W(\bs{s})$ is a random process with unit Pareto margins displaying asymptotic independence. Specifically, the process $W(\bs{s})$ is assumed to satisfy the \citet{Ledford.Tawn:1996} model in \eqref{eq:eta} with coefficient of tail dependence $\eta_W(\bs{s}_1,\bs{s}_2)<1$. In this case, the parameter $\delta$ determines the relative tail heaviness of the terms $R^\delta$ and $W(\bs{s})^{1-\delta}$, and the extent to which one term ``dominates'' the other in the limiting joint tail. Intuitively, when $\delta>0.5$, $R^{\delta}$ is heavier-tailed than $W(\bs{s})^{1-\delta}$ and this yields asymptotic dependence. By contrast, when $\delta<0.5$ $R^{\delta}$ is lighter-tailed than $W(\bs{s})^{1-\delta}$ and this yields asymptotic independence. Formally, \citet{Huser.Wadsworth:2019} showed that we indeed get asymptotic independence for $\delta\leq0.5$ and asymptotic dependence when $\delta>0.5$, and that the coefficient of tail dependence is 
$$\eta(\bs{s}_1,\bs{s}_2)=\begin{cases}
1, & \delta\geq 1/2,\\
\delta/(1-\delta), & \eta_W(\bs{s}_1,\bs{s}_2)/\{1+\eta_W(\bs{s}_1,\bs{s}_2)\} < \delta < 1/2,\\
\eta_W(\bs{s}_1,\bs{s}_2), & \text{otherwise},
\end{cases}$$
{whilst $\chi(\bs{s}_1,\bs{s}_2)=(2\delta-1)\delta^{-1} \E\left[\min\{W(\bs{s}_1),W(\bs{s}_2)\}^{(1-\delta)/\delta}\right]$ when $\delta\geq1/2$.} Thus, the transition between asymptotic dependence classes takes place at $\delta=1/2$, and the strength of extremal dependence interpolates between that of the $W$ process as $\delta\to0$ and perfect dependence as $\delta\to1$. For practical convenience, the $W$ process is typically chosen as a Gaussian process marginally transformed to the unit Pareto scale. Censored likelihood inference for peaks-over-threshold may be performed similarly to ``classical'' Gaussian scale mixtures; see \citet{Huser.Wadsworth:2019}. 
See also \citet{Wadsworth.etal:2017} for a related bivariate model bridging asymptotic dependence classes.

Although the models described in this subsection are quite flexible in their joint tail structures, their main limitation is that the random variable $R$ (or $\tilde{R}$) is constant over space, which prevents them from capturing complete independence as the spatial distance $h=\|\bs{s}_1-\bs{s}_2\|$ increases to infinity. Therefore, these models are usually only realistic over rather small spatial domains, but may not be so over large areas. A related drawback is that, unlike the heavily-parametrized max-mixture models in \eqref{eq:maxmixture}, they cannot capture a change in asymptotic dependence class with distance between sites. More precisely, the asymptotic dependence class is in fact the same for all pairs of sites. Relaxing these limitations is currently an active area of research, and in Section~\ref{sec:Conditional}, we present one recent modeling approach that circumvents these issues by conditioning on single sites being large. Another open research area is the extension to location-scale mixtures, whose tail dependence structures are mostly unknown, though some results exist for special cases such as the skew-$t$ process \citep{Azzalini.Capitanio:2003,Morris.etal:2017,Hazra.etal:2019b} or the generalized hyperbolic distribution originally introduced by \citet{Barndorff-Nielsen:1977,Barndorff-Nielsen:1978}.

\subsection{Max-infinitely divisible processes}\label{subsec:MaxID}
While inverted max-stable, max-mixture, and random scale or location mixture models discussed in Sections~\ref{subsec:invertedMS}--\ref{subsec:RandomScale} are designed to be fitted to peaks over high thresholds, we conclude this section by briefly presenting recent models designed for block maxima, which extend the class of max-stable processes to capture asymptotic independence.

For self-consistency when modeling block maxima, it is natural to restrict ourselves to the class of \emph{max-infinitely divisible} (max-id) distributions. A $D$-dimensional max-id distribution $G$ is such that $G^t$ is a valid distribution for all positive reals $t>0$, and as such, by taking $t=1/m$ for any $m=1,2,\ldots$, they can be seen as the distribution of componentwise maxima over blocks of $m$ random vectors. All univariate distributions are max-id, but this is not the case in the multivariate case $D>1$ (e.g., negatively associated random vectors are not max-id). Moreover, from \eqref{eq:maxstabilityD}, it is evident that max-stable distributions are max-id and are further constrained such that $G^t$ is in the same location-scale family as $G$. 

A random process $Z(\bs{s})$, $\bs{s}\in\calS$, is called max-id if all its finite-dimensional distributions are max-id. Similarly to max-stable processes, max-id processes can be essentially characterized as pointwise maxima over a potentially infinite number of Poisson points on a suitable functions space; see, e.g., \citet{Resnick:1987}, Chapter~5, \citet{Gine.etal:1990} and \citet{Kabluchko.Schlather:2010} for precise theoretical details. However, general max-id models are not necessarily max-stable, and can accommodate more flexible forms of dependence, including asymptotic independence. This was first exploited by \citet{Padoan:2013} for modeling block maxima with dependence strength weakening with increasing event magnitude. The \citet{Padoan:2013} model has an asymptotic justification based on the limit of a specific type of triangular array constructed from Gaussian process ratios with increasing correlation. However, its dependence structure has a rather fast joint tail decay, and this model does not contain max-stable processes within its parameter space, which makes it rather inconvenient in many applications. Alternatively, \citet{Bopp.etal:2020} and \citet{Huser.etal:2020} 
recently proposed different types of max-id models that can capture asymptotic independence, while keeping a popular max-stable process on the boundary of the parameter space. Moreover, the ``distance'' to max-stability in these models is determined via a parameter, so that departure from max-stability can be assessed from the data or, if desired, controlled by the modeler. This is appealing given the long history and strong theoretical justification of max-stable models, and the very wide class of max-id processes.

We here simply discuss one pedagogical example of max-id process construction proposed by \citet{Huser.etal:2020}, which makes a natural link with the spectral representation of max-stable processes in \eqref{eq:maxstableprocess} and the random scale mixture model based on \eqref{eq:Huseretal}. To relax max-stability, while retaining max-infinite divisibility and simultaneously capturing asymptotic independence, \citet{Huser.etal:2020} proposed to mimic the spectral construction \eqref{eq:maxstableprocess}, but to modify the intensity of the Poisson points $\{R_i\}$ in a sensible way. The heavy-tailedness of the intensity $r^{-2}{\rm d}r$ assumed in \eqref{eq:maxstableprocess}, combined with a (rescaled) Gaussian process $W(\bs{s})$, induces asymptotic dependence. Thus, similarly to the Gaussian scale mixture constructions in Section~\ref{subsec:RandomScale}, \citet{Huser.etal:2020} defined a max-id process by assuming in \eqref{eq:maxstableprocess} that $W(\bs{s})$ is a Gaussian process independent of the Poisson points $\{R_i\}$ with Weibull-tailed mean measure $\kappa((r,+\infty])=r^{-\beta}\exp\{-\gamma(r^\beta-1)/\beta\}$, $\beta>0,\gamma>0$. Similarly to the random scale model \eqref{eq:Huseretal}, such a max-id model is asymptotically independent and converges to the asymptotically dependent extremal-$t$ max-stable model with $\gamma$ degrees of freedom \citep{Opitz:2013a} as $\beta\downarrow0$. The parameter $\beta>0$ thus controls the ``distance'' to the extremal-$t$ max-stable model, and provides extra flexibility for capturing the sub-asymptotic behavior of (finite) block maxima.

Inference for max-id models is essentially similar to max-stable models and may be performed by pairwise likelihood, although it may be even more demanding if uni-dimensional integrals similar to \eqref{eq:RWCDF} have to be computed; see \citet{Huser.etal:2020}. 
The max-id model of \citet{Bopp.etal:2020}, however, is amenable to high-dimensional Bayesian inference thanks to its conditional independence representation.

\section{Conditional spatial extremes model}\label{sec:Conditional}
\subsection{Background} \label{sec:ConditionalBackground}
The models described in Section~\ref{sec:AIAD} offer improved flexibility over max-stable and Pareto processes, and typically reflect the extremal characteristics of environmental processes better at finite levels. However, whilst the class of random scale constructions $X(\bs{s}) = R W(\bs{s})$ leads to many useful models, there are two key drawbacks for application to ``larger'' spatial problems:
\begin{enumerate}[(i)]
 \item The need for censored likelihoods becomes prohibitive for more than approximately $30$ observation locations;
 \item The simple construction means that positive dependence persists throughout the spatial domain $\calS$, i.e., $X(\bs{s}_1)$ and $X(\bs{s}_2)$ do not become independent as the distance $\|\bs{s}_1-\bs{s}_2\|$ increases arbitrarily.  
\end{enumerate}
The first of these is predominantly an issue if the spatial problem is ``large'' in the sense of number of observation locations, or grid cells for model output data; the second is an issue if the problem is ``large'' in the sense of a big spatial domain. In practice, both of these problems may be encountered together.  

The conditional spatial extremes model of \citet{Wadsworth.Tawn:2019} was introduced to address these concerns. {Moreover, in contrast to \citet{Morris.etal:2017} who proposed a skew-$t$ process combined with a random partitioning mechanism to break down long-range dependence, the conditional spatial extremes model allows for very flexible forms of extremal dependence and can naturally capture both asymptotic dependence and independence.} The approach builds upon the so-called \emph{conditional extreme-value model} of \citet{Heffernan.Tawn:2004} and \citet{Heffernan.Resnick:2007}. The conditional extreme-value model characterizes the behaviour of a random vector $\bm{Y} \in \mathbb{R}^D$ given that a single component, $Y_j$, is extreme. By analogy, the spatial conditional extremes model focuses on the characterization of a spatial process $Y(\bs{s})$ given that an extreme is observed at an arbitrary location $\bs{s}_0$. Consequently, the approach has clear connections to $r$-Pareto processes discussed in Section~\ref{subsec:Pareto}, since we have already seen that $r(Y) = Y(\bs{s}_0)$ is a valid risk functional. Indeed, the limiting formulation obtained from the theory of $r$-Pareto processes and conditional spatial extremes is identical under asymptotic dependence. However, the advantage of the conditional approach is that limits for asymptotically independent processes can also be handled in a non-trivial way. This is achieved by considering how the extremes of each element of the process $Y(\bm{s})$ changes with $Y(\bm{s}_0)$, in place of assuming that all components of $Y(\bs{s})$ have a positive probability of being jointly extreme simultaneously, as with Pareto processes. 

Specifically, let $X(\bs{s})$ represent the process $Y(\bs{s})$ after a marginal transformation to an exponential-tailed distribution (e.g., exponential, Gumbel or Laplace). If $Y(\bs{s})$ exhibits asymptotic dependence throughout the domain $\calS$, then
\begin{align}
 X(\bs{s}) - X(\bs{s}_0) \mid X(\bs{s}_0)>u \Dto Z^0(\bs{s}),\qquad u \to \infty, \label{eq:extremalincrement}
\end{align}
where $Z^0(\bs{s})$ is a process satisfying $Z^0(\bs{s}_0) = 0$, but with otherwise non-degenerate marginals. We notice that the distance $\|\bs{s}-\bs{s}_0\|$ does not matter in the sense that the normalization of $X(\bs{s})$ required in \eqref{eq:extremalincrement} for the limit to hold does not depend on $\|\bs{s}-\bs{s}_0\|$. In other words, the dependence throughout the process is so strong that when $X(\bs{s}_0)$ is large, the entire process is of the same order of magnitude, such that a simple difference stabilizes to a non-degenerate process. In contrast, the assumption in \citet{Wadsworth.Tawn:2019} generalizes~\eqref{eq:extremalincrement} to
\begin{align}
 \dfrac{X(\bs{s}) - a_{\bs{s}-\bs{s}_0}\{X(\bs{s}_0)\}}{b_{\bs{s}-\bs{s}_0}\{X(\bs{s}_0)\}} \mid X(\bs{s}_0)>u \Dto Z^0(\bs{s}),\qquad u \to \infty,  \label{eq:spatialce}
\end{align}
for some functions $a_{\bs{s}-\bs{s}_0}(\cdot)$, $b_{\bs{s}-\bs{s}_0}(\cdot)>0$ which can depend on the spatial displacement $\bs{s}-\bs{s}_0$. Convergence~\eqref{eq:spatialce} is in the sense of finite-dimensional distributions. 
Under asymptotic dependence, an appropriate choice is $a_{\bs{s}-\bs{s}_0}(x) = x$, $b_{\bs{s}-\bs{s}_0}(x) = 1$, which leads to~\eqref{eq:extremalincrement}. However, under asymptotic independence it is often possible to find $a_{\bs{s}-\bs{s}_0}(\cdot)$ and $b_{\bs{s}-\bs{s}_0}(\cdot)$ that do depend on $\bm{s}-\bm{s}_0$ such that limit~\eqref{eq:spatialce} holds, where limit~\eqref{eq:extremalincrement} would fail. As an example, for the stationary Gaussian process with correlation function $\rho(\cdot)$,
\begin{align*}
 a_{\bs{s}-\bs{s}_0}(x) &= \rho(\bs{s}-\bs{s}_0)^2 x, & b_{\bs{s}-\bs{s}_0}(x) &= 1 +  a_{\bs{s}-\bs{s}_0}(x)^{1/2},
\end{align*}
leads to a limit process $Z^0(\bs{s})$ which is Gaussian and whose correlation structure given in \citet{Wadsworth.Tawn:2019}. The key element to obtaining this non-degeneracy is allowing the normalization of each element of $X(\bs{s})$ to depend on the distance from $X(\bs{s}_0)$.

We note that in many applications, there is no natural conditioning site $\bs{s}_0$. Within the framework of $r$-Pareto processes, one could simply switch to a different risk functional, which will also lead to a non-degenerate formulation due to the asymptotic dependence between all locations. Within the conditional framework, the act of conditioning upon the value at a single location is what leads to the formulation of appropriate models for asymptotically independent processes. \citet{Wadsworth.Tawn:2019} overcome this apparent limitation by using a composite likelihood to combine information and introducing an importance sampling scheme to change the conditioning event {to the process being extreme at any of one of an arbitrary set of locations, i.e., $\max_{1\leq j\leq D} X(\bs{s}_j)>v$ for some large threshold $v$. See also \citet{Gilleland.etal:2013} for a related (non-parametric) spatial modeling approach applied to climatology, where conditioning is performed upon a spatial functional of the process (rather than a single point) being large.}

\subsection{Model}
\label{sec:ConditionalModel}

To translate limit~\eqref{eq:spatialce} into a statistical model, \citet{Wadsworth.Tawn:2019} assume that for $X(\bs{s}_0)>u$, with $u$ a high marginal threshold,
\begin{align}
 X(\bs{s}) \mid X(\bs{s}_0)>u \approx a_{\bs{s}-\bs{s}_0}\{X(\bs{s}_0)\} + b_{\bs{s}-\bs{s}_0}\{X(\bs{s}_0)\} Z^0(\bs{s}), \label{eq:conditionalmodel}
\end{align}
where the specification of functions $a_{\bs{s}-\bs{s}_0},b_{\bs{s}-\bs{s}_0}$ and distribution of the process $Z^0$ complete the specification of the model. The additional aspect over using a Pareto process model is that $a_{\bs{s}-\bs{s}_0}$ and $b_{\bs{s}-\bs{s}_0}$ are chosen as part of the model rather than prespecified. Based on a range of theoretical examples, and desirable model properties, they consider the form $a_{\bs{s}-\bs{s}_0}(x) = \alpha(\bs{s}-\bs{s}_0) x$ with
\begin{align}
\label{eq:alpha}
 \alpha(\bs{s}-\bs{s}_0) &= \begin{cases}
                             1, & \|\bs{s}-\bs{s}_0\| \leq \Delta\\
                             \exp\{-(\|\bs{s}-\bs{s}_0\|-\Delta)^\kappa/\lambda\}, & \|\bs{s}-\bs{s}_0\|> \Delta,
                            \end{cases}
\end{align}
where $\Delta\geq0$, $\lambda>0$ and $\kappa>0$. The rationale for such a choice is that it allows modeling of asymptotic dependence up to some spatial displacement $\Delta$, and asymptotic independence with weakening dependence beyond this lag. Other functional forms could be used for $\|\bs{s}-\bs{s}_0\|> \Delta$, such as alternative correlation or survival functions. The fact that the form in equation~\eqref{eq:alpha} links only to distance $\|\bs{s}-\bs{s}_0\|$, rather than direction or location is reasonable under an assumption that the process being modeled is stationary and isotropic. At the end of this section, we discuss possible approaches to handle non-stationarity.

Three different forms for $b_{\bs{s}-\bs{s}_0}$ were considered by \citet{Wadsworth.Tawn:2019} to achieve different modeling aims. We detail only one of these, which in conjunction with an appropriate form for $Z^0$, permits independence of $X(\bs{s})$ and $X(\bs{s}_0)$ when $\|\bs{s}-\bs{s}_0\|$ is sufficiently large. Specifically, this can be achieved by $b_{\bs{s}-\bs{s}_0}(x) = 1 + a_{\bs{s}-\bs{s}_0}(x)^\beta$, $\beta \in [0,1)$, since for large $\|\bs{s}-\bs{s}_0\|$ we then have $a_{\bs{s}-\bs{s}_0}(x) \approx 0$,  $b_{\bs{s}-\bs{s}_0}(x) \approx 1$ and hence $X(\bs{s}) \mid X(\bs{s}_0)>u \approx Z^0(\bs{s})$. To complete this specification, the marginal distribution of $Z^0(\bs{s})$ should be the same as that of $X(\bs{s})$. This may be handled by taking the marginal distributions of $X$ as Laplace, and specifying the margins of $Z^0$ to have density
\begin{align}
 f(z) = {\delta\over 2\sigma\Gamma(1/\delta)}\exp\{-\left|{(z-\mu)/\sigma}\right|^{\delta}\}, \qquad \delta>0, \label{eq:deltaLaplace}
\end{align}
which includes the Gaussian and Laplace densities as special cases for $\delta=2$ and $\delta=1$, respectively. To achieve approximate independence with increasing distance in model~\eqref{eq:conditionalmodel}, the parameters in~\eqref{eq:deltaLaplace} should evolve with $\|\bs{s}-\bs{s}_0\|$ such that when this quantity is large, $\mu(\|\bs{s}-\bs{s}_0\|) \approx 0$, $\sigma(\|\bs{s}-\bs{s}_0\|) \approx 1$ and $\delta(\|\bs{s}-\bs{s}_0\|) \approx 1$. If independence is not observed over the size of the domain, then these restrictions need not apply. 

The model is completed by assuming a Gaussian process dependence structure for $Z^0$ which makes for simpler likelihoods and hence permits inference in moderately high dimensions. To ensure the constraint that $Z^0(\bs{s}_0) = 0$, one can begin with a Gaussian process $Z_G(\bs{s})$ and either take $Z_G(\bs{s}) - Z_G(\bs{s}_0)$ or $Z_G(\bs{s})\mid Z_G(\bs{s}_0)=0$, which yield new Gaussian processes with the desired property. From there, marginal transformations can be applied as desired. For example, the parameters $\mu(\|\bs{s}-\bs{s}_0\|)$, $\sigma(\|\bs{s}-\bs{s}_0\|)$ could follow a structure implied by a Gaussian process specification or be parameterized independently, as with $\delta(\|\bs{s}-\bs{s}_0\|)$. An example parameterization of the latter is $\delta(\|\bs{s}-\bs{s}_0\|) = 1 + \exp\{-(\|\bs{s}-\bs{s}_0\|/\delta_1)^{\delta_2}\}$.

A simpler version of the model described in this section has been used by \citet{Shooter.etal:2019} to model hindcast significant wave height data on one-dimensional transects in the North Sea.

When undertaking inference for extremes over a large spatial domain, the assumptions of stationarity and isotropy over the domain become less {plausible}. We focus on the more difficult problem of nonstationarity, since anisotropy can usually be remedied by including a suitable linear coordinate transformation into the inference \citep[see, e.g.,][]{Blanchet.Davison:2011,Huser.etal:2017}. In the context of max-stable processes, \citet{Huser.Genton:2016} proposed the use of non-stationary covariance functions in the Gaussian processes that formed part of the max-stable spectral representation. A similar approach could be taken here if covariates are available. Furthermore, these covariates could enter into any aspect of the model: $a_{\bs{s}-\bs{s}_0}$, $b_{\bs{s}-\bs{s}_0}$ or the covariance structure of $Z^0$; \citet{Jonathan.etal:2014} present related ideas in the multivariate context. An alternative approach taken by \citet{Cooley.etal:2007} is to consider so-called ``climate space" coordinates of proxies that are related to the observations, rather than geographic coordinates, but again this requires knowledge of relevant variables. When covariates are not available, a sensible alternative is the spatial deformation approach first introduced by \citet{Sampson.Guttorp:1992}, {though it would ideally need to be tailored to extremal dependence.} 


\subsection{Inference}
\label{sec:ConditionalInference}
Conditioning only upon the process at a single site, $\bs{s}_0$, being large leads straightforwardly to a likelihood for inference, by combining equation~\eqref{eq:conditionalmodel} with the specifications for $a_{\bs{s}-\bs{s}_0}$, $b_{\bs{s}-\bs{s}_0}$ and $Z^0$. However, under an assumption of stationarity, the model parameters have the same form regardless of conditioning site. Consequently, for $D$ observation locations, \citet{Wadsworth.Tawn:2019} propose to combine the resulting $D$ likelihoods by multiplying them to form one composite likelihood; see \citet{Varin.etal:2011} for an overview of composite likelihoods. By maximizing this composite likelihood, a single set of parameter estimates is obtained, which should, on average, represent the process well at all locations. Assessment of parameter uncertainty may be undertaken by nonparametric (block) bootstrap. 

We note that the likelihood is composite because of the fact that processes $X$ with ${X(\bs{s}_j)>u}$ at more than one site $\bs{s}_j$ will appear multiple times in the likelihood due to the different conditioning sites. As a consequence, composite likelihood inference takes longer than selecting a single conditioning site, but parameter estimates are not too tailored to any one location. A compromise for large $D$ is to combine over a subset of $D' \ll D$ of conditioning sites; this is implemented in the  application to Irish temperature extremes, presented in Section~\ref{subsec:irish}. Nonetheless, an advantage over existing methodology is that censored likelihoods are not necessary, because the conditional extremes methodology is tailored to allow for moderate and small values occurring alongside large values.

\section{Environmental applications}\label{sec:applications}

\subsection{Dutch wind speed data}\label{subsec:wind}
\begin{figure}[t!]
\centering
 \includegraphics[width=0.49\textwidth]{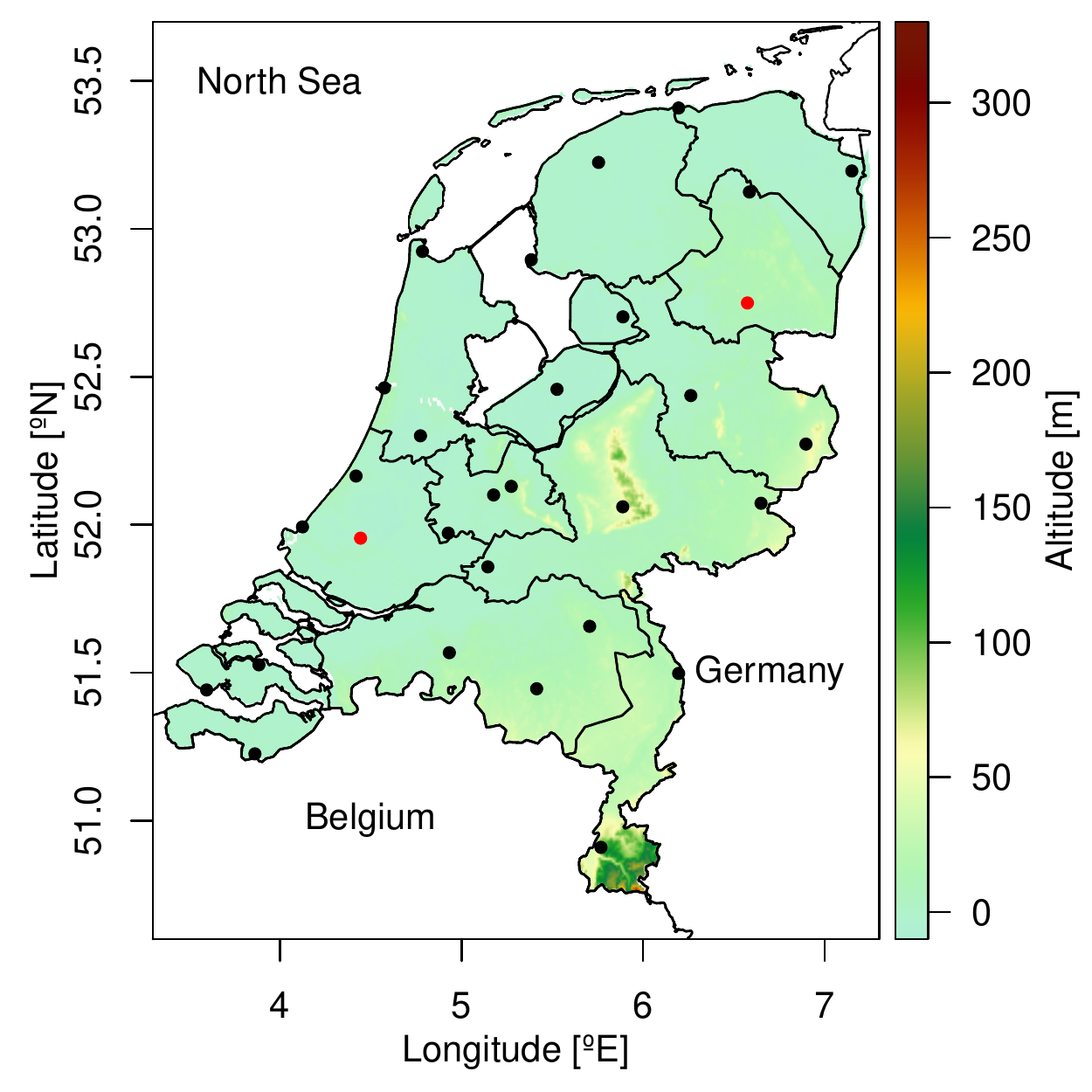}
 \includegraphics[width=0.49\textwidth]{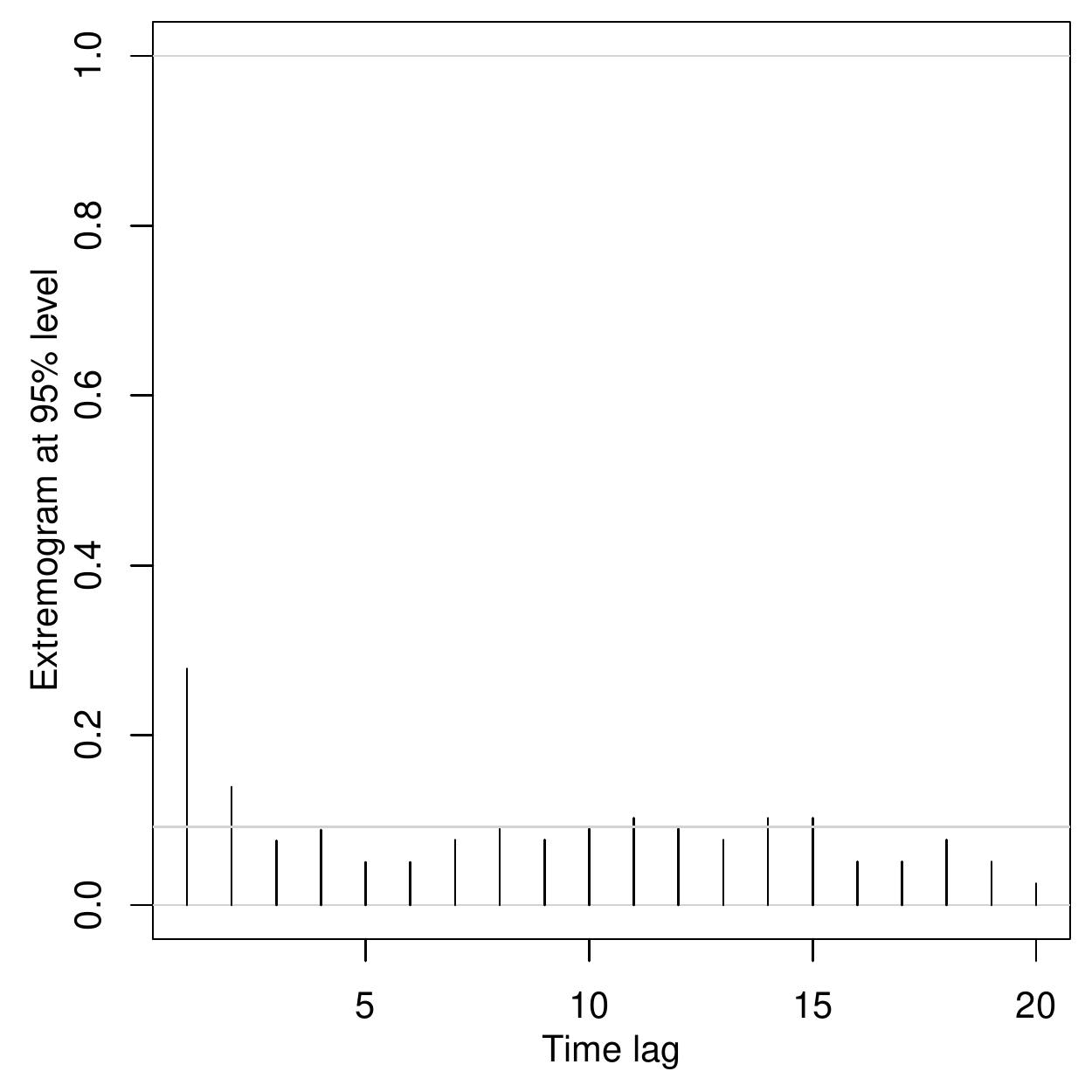}
  \caption{Left: Topographic map of the Netherlands (study region) and neighboring countries, with monitoring stations indicated by dots. Red dots are stations selected to display model diagnostics in Figure~\ref{fig:diag}. Right: Extremogram ${\pr\{U_{t+h}(\bs{s}_{14})>0.95\mid U_{t}(\bs{s}_{14})>0.95\}}$ plotted against time lag $h=1,2,\ldots,20$, for the $14$-th station $\bs{s}_{14}$ (with coordinates $6.575^\circ$E, $52.75^\circ$N, shown in red on the left panel). The horizontal grey line is a bootstrap $95\%$ upper confidence bound under independence.}
 \label{fig:map}
\end{figure}

In our first application, we compare the performance of some asymptotic $r$-Pareto (Section~\ref{subsec:Pareto}) and sub-asymptotic random scale mixture models (Section~\ref{subsec:RandomScale}) by re-analyzing the Dutch wind speed data studied by \citet{Opitz:2016} and \citet{Huser.etal:2020} among others. Evidence of asymptotic independence was found in these papers, either based on threshold exceedances or block maxima, respectively. The dataset comprises daily wind speed measurements from December 24, 1999, to November 16, 2008, at $D=30$ stations spread across the Netherlands; see the left panel of Figure~\ref{fig:map}. Latitude--Longitude coordinates were transformed first to a metric system to avoid issues of interpretation. To avoid the modeling of seasonality, we restrict ourselves to the months of October to March, when the strongest wind speeds usually occur. Over this period, only 10 days (i.e., 10 days $\times$ 30 stations $=300$ observations) are missing, which yields $n=1594$ non-missing days in total. Let $Y_t(\bs{s}_j)$ denote the observed process at time $t=1,\ldots,n$, and station $\bs{s}_j\in\calS\subset\Real^2$, $j=1,\ldots,D$, where $\calS$ represents the study region. For simplicity, we here standardize the data at each station separately using the empirical distribution function as $U_t(\bs{s}_j)={\rm rank}\{Y_t(\bs{s}_j)\}/(n+1)$, where the rank is taken over the $n$ observations at each station, {and we here ignore the presence of ties by randomizing their ranks. Although such an approach is quite common in the literature, more sophisticated methods to handle ties would in principle be desirable.} To assess extremal dependence in time, we compute the extremogram ${\pr\{U_{t+h}(\bs{s}_j)>u\mid U_{t}(\bs{s}_j)>u\}}$ at lags $h=1,2,\ldots$, for some high threshold $u\in(0,1)$ at each station $\bs{s}_j$. The right panel of Figure~\ref{fig:map} shows the results for $u=0.95$ and station $\bs{s}_{14}$. Some weak extremal dependence exists at lag 1, but it rapidly vanishes at larger lags and as the threshold $u$ increases (not shown). Similar results hold for other stations. In the following, we fit various models by censored likelihood, treating days as independent time replicates.

As there is evidence of geometric anisotropy in the data, we first fit the {extremal} dependence structure of a stationary Gaussian process with powered exponential correlation function 
\begin{equation}\label{eq:correlation}
\rho(\bs{s}_1,\bs{s}_2)=\exp\left\{-(\sqrt{(\boldsymbol{s}_1-\boldsymbol{s}_2)^T\Omega^{-1}(\boldsymbol{s}_1-\boldsymbol{s}_2)}/\phi)^\nu\right\},
\end{equation}
where $\phi>0$ and $\nu\in(0,2)$ are global range and smoothness parameters, respectively, and 
\begin{equation}\label{eq:aniso}
\Omega=\begin{pmatrix}
\cos(\psi) & -\sin(\psi)\\
\sin(\psi) & \cos(\psi)
\end{pmatrix}\begin{pmatrix}
1 & 0\\
0 & L^{-2}
\end{pmatrix}\begin{pmatrix}
\cos(\psi) & -\sin(\psi)\\
\sin(\psi) & \cos(\psi)\\
\end{pmatrix}^T,
\end{equation}
with $\psi\in(-\pi/2,\pi/2)$ and $L>0$ the rotation and stretch parameters. When $L=1$, the model becomes isotropic. To fit this Gaussian copula model to threshold exceedances, we use a likelihood function that censors observations below the $95\%$ marginal level. The estimated anisotropy parameters (standard errors) are $\hat\psi=-1.08$ ($0.06$) and $\hat L=0.53$ ($0.04$), indicating slightly stronger spatial dependence along the coast than in the perpendicular direction. We then plug these anisotropy parameter estimates into \eqref{eq:aniso} to get $\hat\Omega=\hat\Omega^{1/2}\hat\Omega^{T/2}$, and fit various isotropic extremal dependence models based on a modified set of stations defined through the linear transformation
\begin{align}\label{eq:transformation}
\bs{s}_j^\star=\hat\Omega^{-1/2}\bs{s}_j=
\begin{pmatrix}
1 & 0\\
0 & \hat L
\end{pmatrix}\begin{pmatrix}
\cos(\hat \psi) & \sin(\hat \psi)\\
-\sin(\hat \psi) & \cos(\hat \psi) 
\end{pmatrix}\bs{s}_j,\qquad j=1,\ldots,D.
\end{align}
{Although the geometric anisotropy parameters could in principle be estimated jointly with the other parameters in all models considered below, estimating them in a preliminary step offers a significant speed-up and should only very slightly impact our model comparison.}

Specifically, we fit the \citet{Huser.etal:2017} Gaussian scale mixture \eqref{eq:Huseretal} with $\beta>0$ and $\gamma>0$, as well as the limit model obtained as $\beta\downarrow0$. Recall that the model with $\beta>0$ leads to asymptotic independence, while $\beta\downarrow0$ leads to asymptotic dependence, and resembles the dependence structure of a Student's $t$ process. We also fit the hybrid model of \citet{Huser.Wadsworth:2019} (recall Section~\ref{subsec:RandomScale}), which bridges dependence classes in the interior of the parameter space. To contrast these sub-asymptotic models with more classical asymptotic spatial extreme-value models, we also fit the $r$-Pareto process derived from the risk functional $r(\tilde{Y})=\max_{1\leq j\leq D}\tilde{Y}(\bs{s}_j)$ \citep{deFondeville.Davison:2018}. Finally, for comparison, we also include the Gaussian copula model. For consistency, all Gaussian process components within these models are based on the isotropic powered exponential correlation function $\rho(\bs{s}_1,\bs{s}_2)={\exp\{-(\|\bs{s}_1-\bs{s}_2\|/\phi)^\nu\}}$, with range $\phi>0$ and smoothness $\nu\in(0,2)$, except for the $r$-Pareto process where we use the Brown--Resnick formulation based on variogram $(\|\bs{s}_1-\bs{s}_2\|/\phi)^\nu$, $\phi>0,\nu\in(0,2)$. Moreover, all models are fitted to threshold exceedances based on a censored likelihood using $u=0.95$ as the threshold probability level, though---unlike the other models---the likelihood for the $r$-Pareto process conditions on having at least one threshold exceedance and thus does {not} involve the contribution of observation vectors that are fully censored.

Table~\ref{tab:results} reports the results in terms of estimated parameters, their standard errors calculated based on the {observed information}, the maximized log-likelihood values, and the corresponding Bayesian information criterion (BIC). In all fitted models, the range parameter $\phi$ is quite high indicating rather strong spatial dependence overall, while the smoothness parameter $\nu$ is quite low indicating small-scale variability. The unconstrained \citet{Huser.etal:2017} model has $\hat\beta=2.52$ with standard error $0.31$. This suggests that the data are asymptotically independent and may {be} well described by the Laplace model of \citet{Opitz:2016}, which has a Weibull index of $\beta=2$. However, as lower values of $\gamma$ imply stronger dependence, the estimated value of $\hat\gamma=0.003$ makes it difficult to determine the asymptotic dependence class with high certainty. When $\beta$ is fixed to zero (i.e., $\beta\downarrow0$), we get $\hat\gamma=6.97$ instead. These parameter estimates imply that our dataset is somewhere in between strong asymptotic independence and weak asymptotic dependence. The \citet{Huser.Wadsworth:2019} model, with the transition between extremal dependence classes in the interior of its parameter space, clears any doubt. With this model, we get $\hat\delta=0.44$ with estimated $95\%$ confidence interval about $(0.40,0.48)$. As the critical point of $\delta=0.5$ does not lie within the confidence interval, this indicates that there is quite strong support for asymptotic independence.

\begin{table}[t!]
	\centering
	\caption{Estimated parameters and standard errors (subscripts), maximized log-likelihood values and Bayesian information criterion (BIC) for the different models fitted in the Dutch wind speed application. HOT refers to the \citet{Huser.etal:2017} Gaussian scale mixture model based on \eqref{eq:Huseretal}, while HW refers to the \citet{Huser.Wadsworth:2019} model.} 
	\vspace{10pt}
	\label{tab:results}
	\begin{tabular}{r|ccccc|cc}
		\hline
		Model & $\log \phi$ & $\nu$ & $\beta$ & $\gamma$ & $\delta$ & log-lik. & BIC \\ \hline
		Gaussian & $9.71_{0.28}$ & $0.40_{0.02}$ &  &  &  &  $4242.2$ & $-8469.5$\\ 
		HOT, $\beta\downarrow0$ & $9.38_{0.26}$ & $0.41_{0.02}$ & $0$ & $6.97_{0.43}$ & & $4290.2$ & $-8558.4$\\ 
		HOT, $\beta>0$ & $8.70_{0.28}$ & $0.41_{0.02}$ & $2.52_{0.31}$ & $0.003_{0.005}$ & & $4294.1$ & $-8558.7$\\
		HW & $8.57_{0.09}$ & $0.42_{0.01}$ &  &  & $0.44_{0.02}$ & $4292.7$ & $-8563.4$\\
		$r$-Pareto & $5.62_{0.03}$ & $0.37_{0.01}$ &  &  &  & $4157.7$ & $-8300.6$\\ \hline
	\end{tabular}
\end{table}

\begin{figure}[t!]
\centering
 \includegraphics[width=0.8\textwidth]{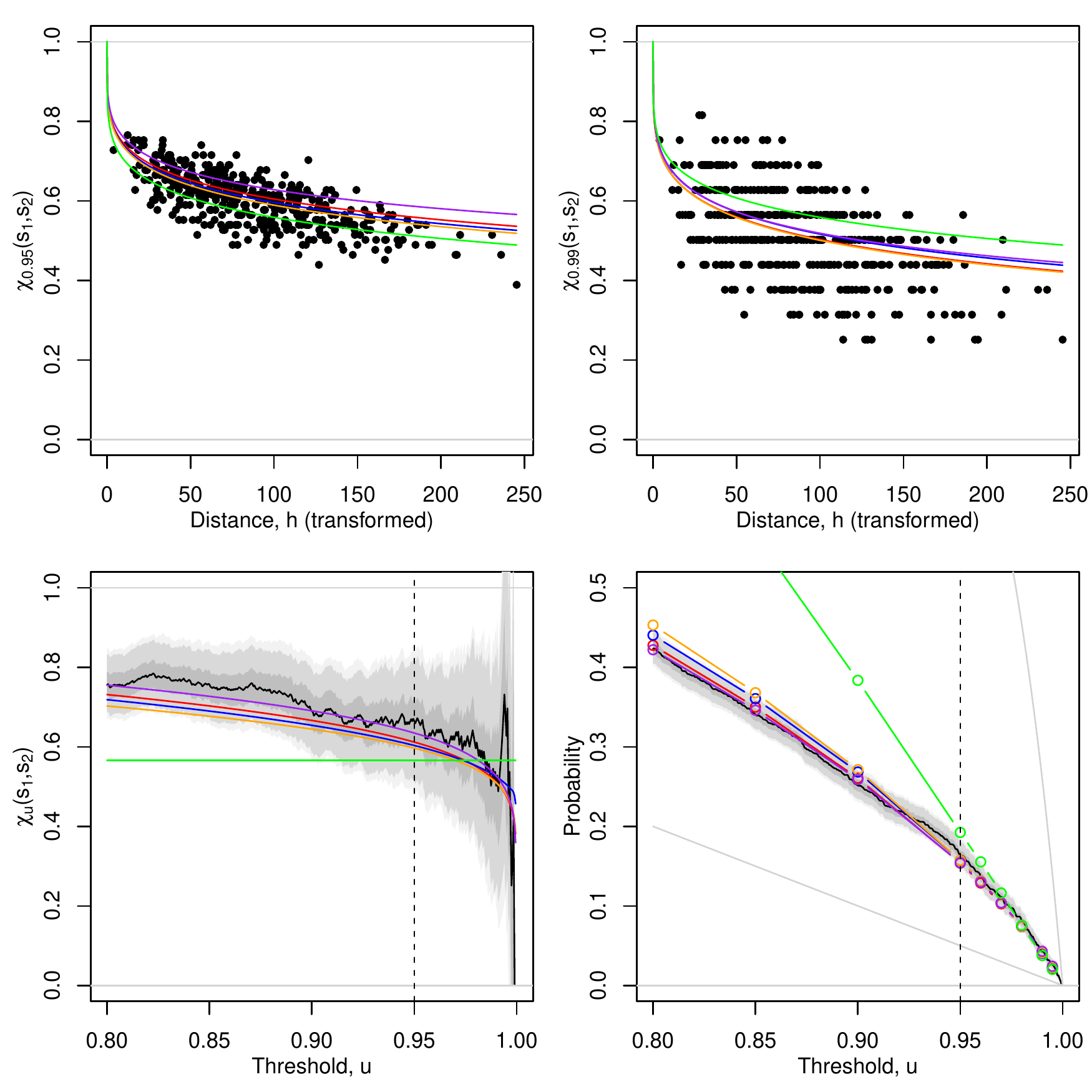}
  \caption{Top: Coefficient $\chi_u(\bs{s}_1,\bs{s}_2)$ plotted for $u=0.95$ (left) and $u=0.99$ (right) against the transformed distance $h=\|\bs{s}_1^\star-\bs{s}_2^\star\|$. Black dots are empirical estimates for all pairs of stations, while solid curves are the Gaussian copula model (red), the \citet{Huser.etal:2017} model with $\beta=0$ (blue) and $\beta>0$ (orange), the \citet{Huser.Wadsworth:2019} model (purple) and the $r$-Pareto process (green). Bottom: Coefficient $\chi_u(\bs{s}_1,\bs{s}_2)$ (left) for a pair of sites at moderate distance from each other (red dots in Figure~\ref{fig:map}), and probability $\pr\{\max_{1\leq j\leq D}U_t(\bs{s}_j)>u\}$ (right), plotted for various thresholds $u\in(0.8,1)$. Black and colored curves are as in the top panels. Gray shaded areas are $50\%,90\%,95\%$ (darker to lighter) pointwise confidence bands for empirical estimates. Vertical dashed lines represent the threshold $u=0.95$ used for fitting using the censored likelihood approach.}
 \label{fig:diag}
\end{figure}

According to the BIC values, the best model overall is the \citet{Huser.Wadsworth:2019} model, although the Gaussian scale mixture model of \citet{Huser.etal:2017} has a quite similar performance. These models, however, show a major improvement with respect to the Gaussian copula model, which is asymptotically independent but too rigid in its tail decay rate, and the (asymptotically dependent) $r$-Pareto model, which is unable to capture weakening dependence at increasingly high quantile level. With its substantially larger BIC value (lower log-likelihood), the $r$-Pareto process is in fact even worse than the Gaussian copula model. {However, care is needed when using the BIC to compare the $r$-Pareto process (whose likelihood conditions on exceeding the threshold in at least one location) to the other models. Unreported calculations show that the gap in log-likelihoods and BIC values between the $r$-Pareto process and the other models is even \emph{larger} than it appears in Table~1 when all models are restricted to the support of the $r$-Pareto process. Precisely, when the fitted models are all compared after conditioning upon an extreme event in at least one location, the censored log-likelihoods of the Gaussian copula, \citet{Huser.etal:2017}, and \citet{Huser.Wadsworth:2019} models indeed increase by about 715, which makes them much more attractive than the $r$-Pareto process for this particular dataset.} Figure~\ref{fig:diag} illustrates the goodness-of-fit of the different models through visual diagnostics. All models seem to perform decently well overall, but again, the $r$-Pareto process tends to largely overestimate the dependence strength at high quantiles (and underestimate it at lower quantiles), owing to its threshold-stability property \eqref{eq:threshstab}.

\subsection{Irish temperature data}\label{subsec:irish}

To illustrate the spatial conditional extremes model described in Section~\ref{sec:Conditional}, we now fit it to a dataset of daily maximum summer temperatures from Ireland and Northern Ireland. The values comprise a subset of the E-OBS dataset\footnote{Data available from: \texttt{http://surfobs.climate.copernicus.eu/dataaccess/access\_eobs.php}} of daily maximum temperatures on a 0.25$^\circ$ grid, and we focus on the 178 grid locations covering the island of Ireland during the summer months (June, July and August) of the 16 year-period 1995--2010.

\begin{figure}[t!]
\centering
 \includegraphics[width=0.49\textwidth]{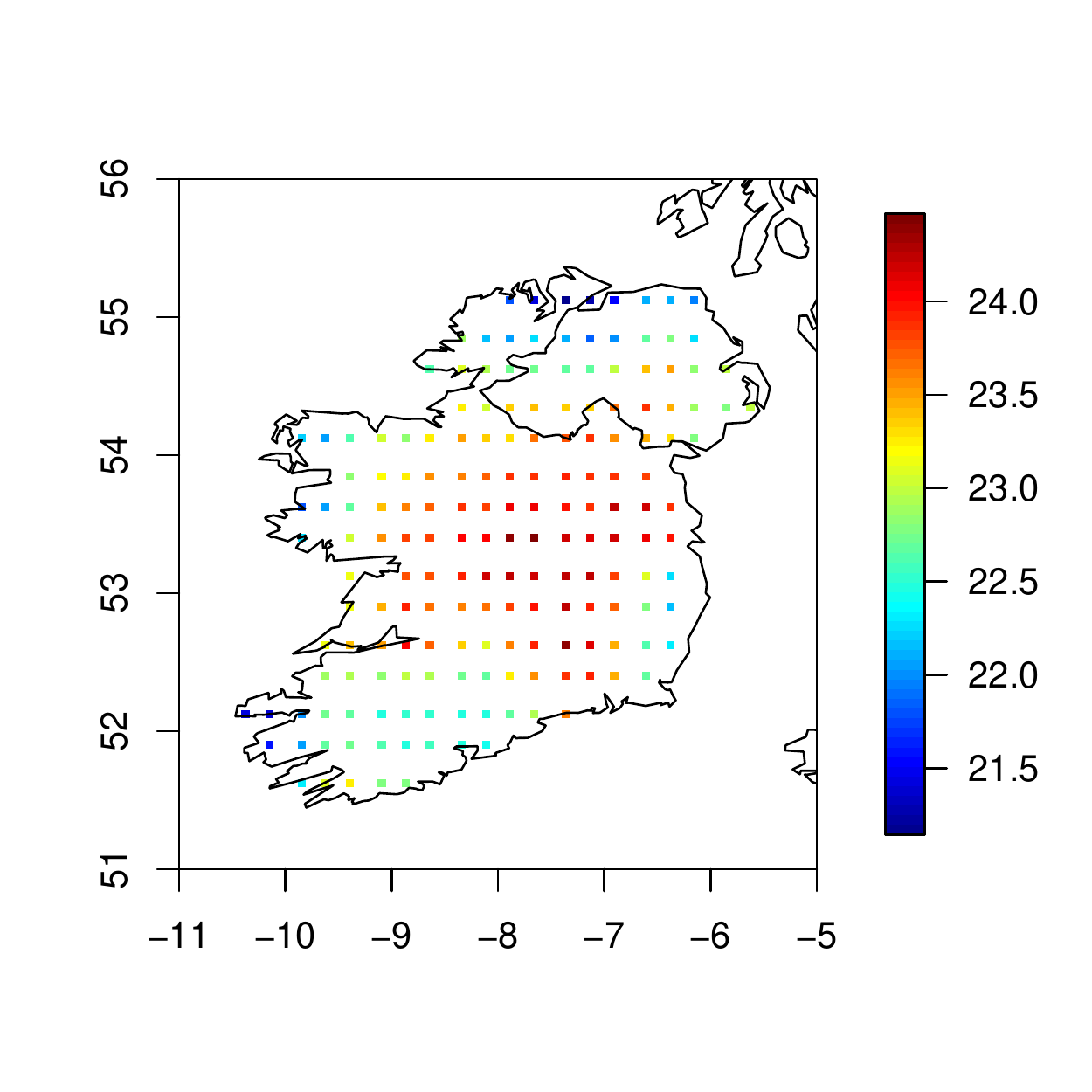}
 \includegraphics[width=0.49\textwidth]{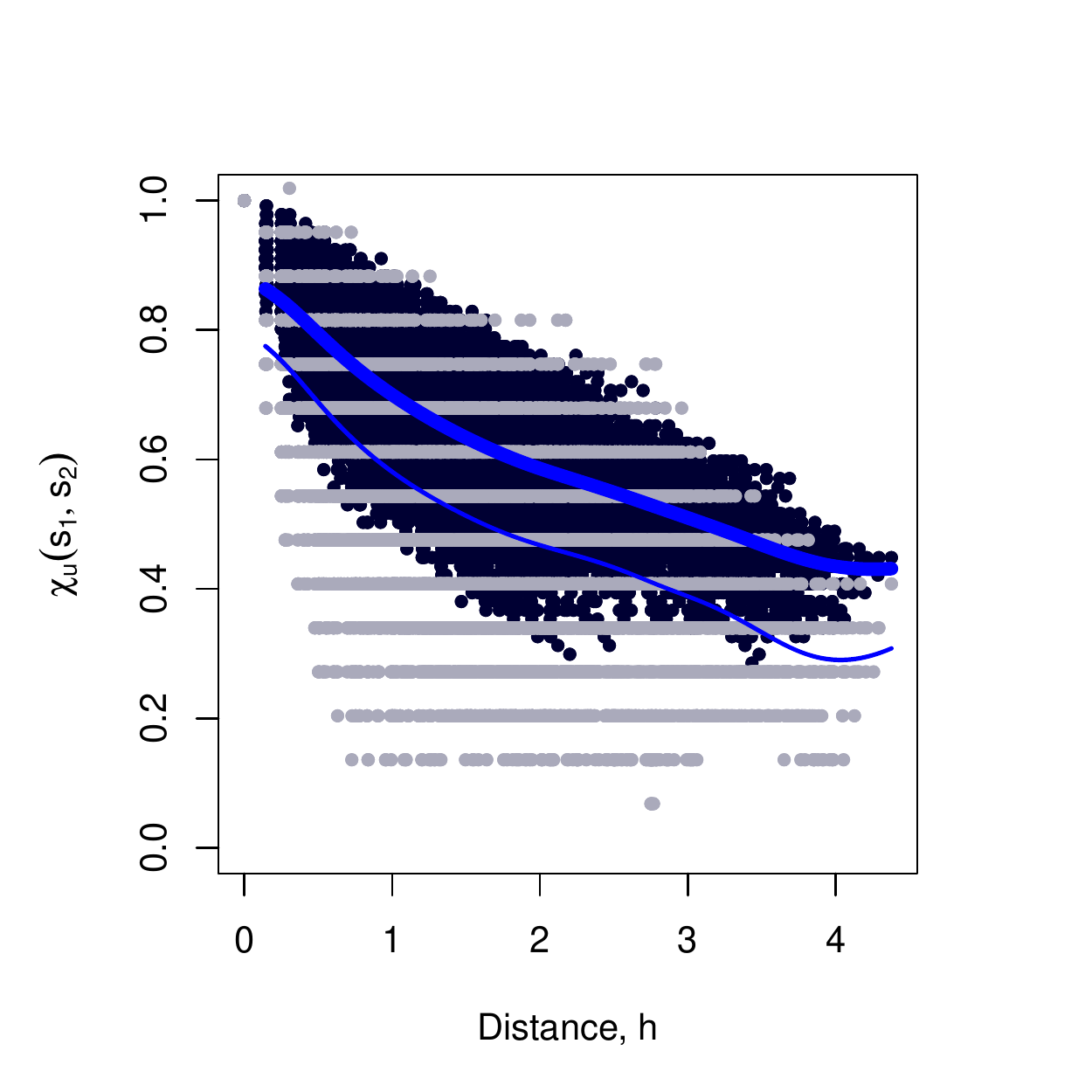}
  \caption{Left: Marginal 95\% quantiles at the 178 grid locations. Right: estimates of $\chi_{0.95}(\bs{s}_1,\bs{s}_2)$ (dark points) and $\chi_{0.99}(\bs{s}_1,\bs{s}_2)$ (light points) against distance $h = \|\bs{s}_1-\bs{s}_2\|$ in units of latitude. Lines represent kernel smoothed estimates at $u=0.95$ (thick line) and $u=0.99$ (thin line).}
 \label{fig:IrelandExploratory}
\end{figure}

The data were transformed to have approximately Laplace marginals by using the empirical distribution function at each site. Site-wise transformation accounts for marginal non-stationarity, which occurs in most spatial datasets and is evident from the empirical 95\% quantiles displayed in Figure~\ref{fig:IrelandExploratory}.

The coordinates of the data were transformed such that units of longitude are approximately equal to units of latitude (with one unit $\approx 111$km). Figure~\ref{fig:IrelandExploratory} displays estimates of $\chi_u(\bs{s}_1,\bs{s}_2)$ against distance $h=\|\bs{s}_1-\bs{s}_2\|$ and evaluated at different uniform quantiles $u\in(0,1)$. The estimates suggest that positive extremal dependence persists over the whole domain, and that there is some decrease in the strength of the dependence at higher quantiles. This could be consistent either with an asymptotically independent model, or a sub-asymptotic asymptotically dependent model. Either of these are possible within the conditional framework, but in contrast to the model of \citet{Huser.Wadsworth:2019}, there is a need to select the model manually using likelihood values and goodness-of-fit diagnostics. 

Overall, accounting for different possibilities in the functions $a_{\bs{s}-\bs{s}_0},b_{\bs{s}-\bs{s}_0}$ and the residual process $Z^0$, there are a large number of potential models. For brevity, following some preliminary investigation, we focus on Model~2 from \citet{Wadsworth.Tawn:2019}, with $a_{\bs{s}-\bs{s}_0}(x) = \alpha(\|\bs{s}-\bs{s}_0\|) x$ and $\Delta=0$ as described in~\eqref{eq:alpha} and $b_{\bs{s}-\bs{s}_0}(x)= x^\beta$. The residual process is taken as $Z^0(\bs{s}) = t_{\bs{s}-\bs{s}_0}(\tilde{Z}^0(s))$, where $\tilde{Z}^0(s)$ has the same distribution as $Z_G(\bs{s})\mid Z_G(\bs{s}_0) = 0$ for $Z_G(\bs{s}_0)$ a stationary Gaussian process with mean $\mu$ and covariance function $C(\bs{s}_1,\bs{s}_2) = \sigma^2 \rho(\bs{s}_1,\bs{s}_2)$ with $\sigma>0$ and anisotropic correlation $\rho(\bs{s}_1,\bs{s}_2)$ given as in \eqref{eq:correlation}, and $t_{\bs{s}-\bs{s}_0}(\cdot)$ maps the marginals of the conditional Gaussian process to those with density~\eqref{eq:deltaLaplace}. In total, the model has 10 parameters.

We use the composite likelihood approach to inference as described in Section~\ref{sec:ConditionalInference}, but involving only a subset $D'=30$ of the possible $D=178$ conditioning sites. We emphasize that information from each of the 178 sites is still being included in this approach, but that using a subset of conditioning sites reduces the burden for high dimensions whilst maintaining the general principle that the composite likelihood helps to achieve a single set of parameter estimates that on average represent the process well. Each component of the composite likelihood is the full conditional likelihood conditioning upon $X(\bs{s}_j)>u$, with $u$ the $95\%$ quantile of the Laplace distribution. Parameter estimates and summaries of the bootstrap distribution are displayed in Table~\ref{tab:IrelandResults}. The parameters $\kappa$ and $\lambda$ relate to the function $\alpha(\|\bs{s}-\bs{s}_0\|)$ in~\eqref{eq:alpha}; $\beta$ controls $b_{\bs{s}-\bs{s}_0}(x)$; $\phi,\nu,\sigma$ control the covariance function of $Z(\bs{s})$; $\mu$ represents the mean of $Z_G(\bs{s})$, from which $Z^0(\bs{s})$ is derived. The quantity $\delta$ is the shape parameter of the delta-Laplace density in~\eqref{eq:deltaLaplace}, while the location and scale parameters of that density are obtained by matching those from $\tilde{Z}^0(\bs{s})$. Finally, $\psi$ and $L$ represent the rotation and stretch parameters for the geometric anisotropy, defined as in \eqref{eq:aniso}.

\begin{figure}[t!]
\centering
 \includegraphics[width=0.45\textwidth]{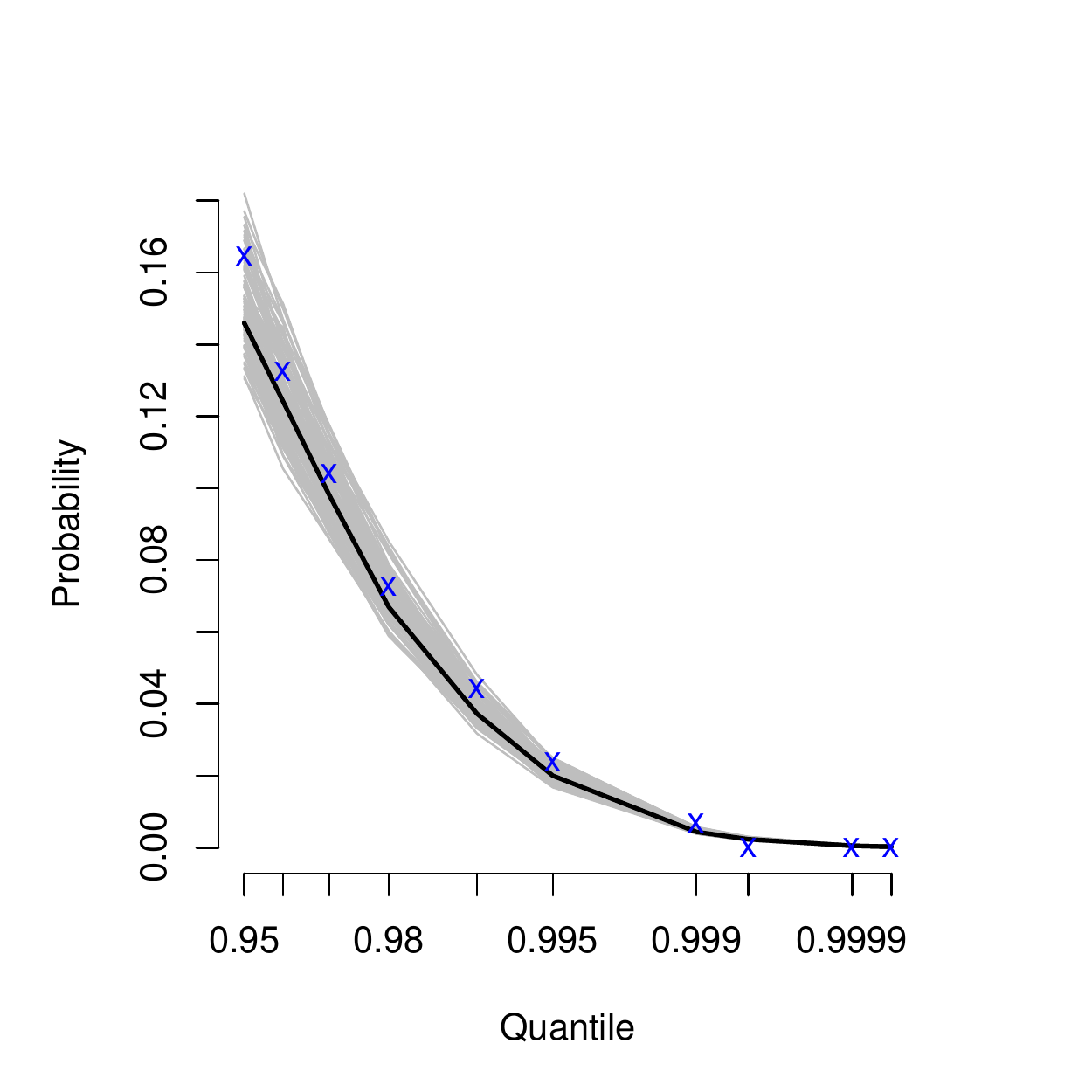}
\includegraphics[width=0.45\textwidth]{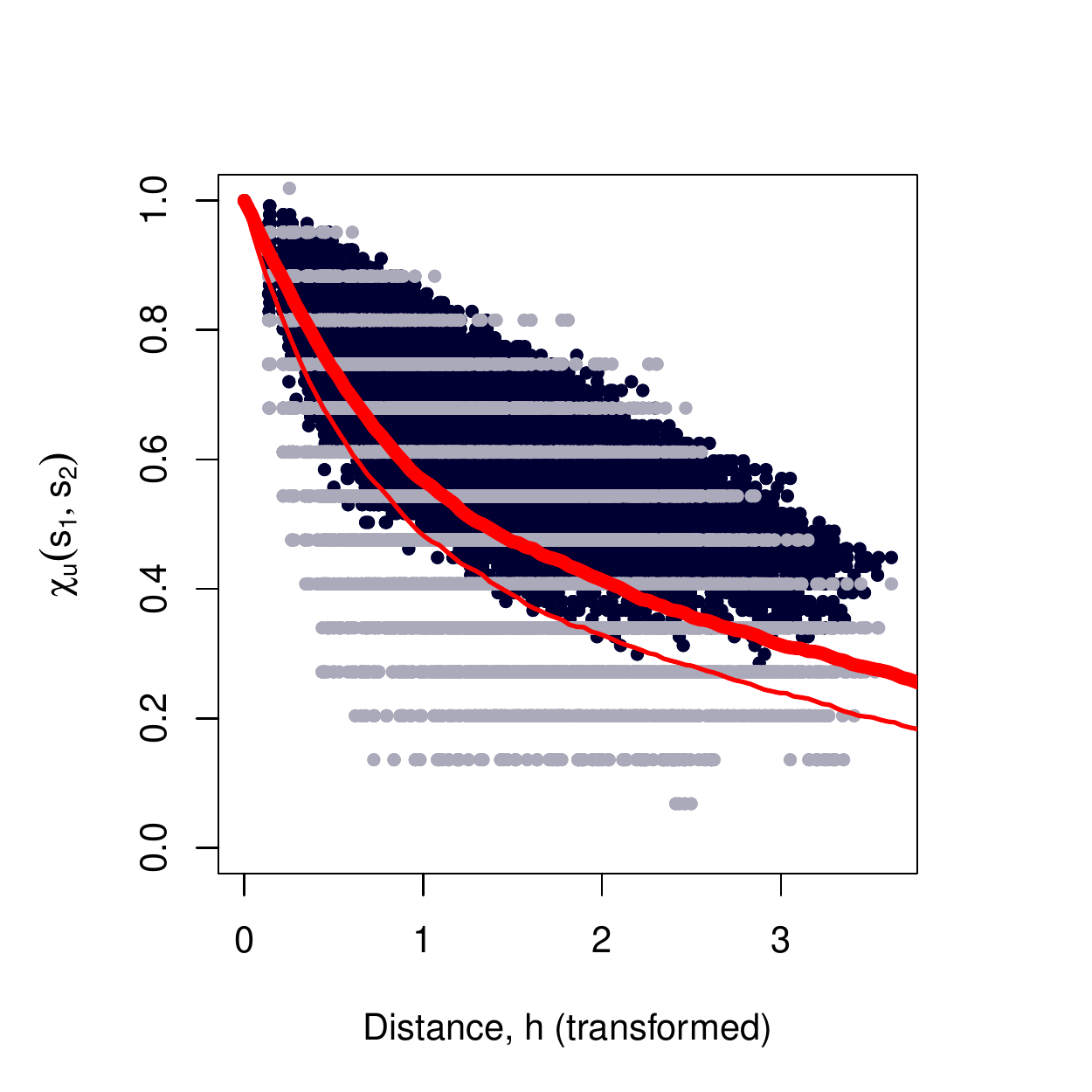} \caption{Left: Model-based estimate of $\Pr\{\max_{1 \leq j \leq D} X(\bs{s}_j) > v\}$ (solid line, estimates based on bootstrapped parameter values in grey), and empirical values from the data (blue crosses). Right: estimates of $\chi_{0.95}(\bs{s}_1,\bs{s}_2)$ (darker points) and $\chi_{0.99}(\bs{s}_1,\bs{s}_2)$ (lighter points) against distance in coordinates transformed to account for anisotropy (dots), and estimates from fitted model (red lines). }
 \label{fig:IrelandResults}
\end{figure}

The estimates show that over the spatial range of the island, $\widehat{\alpha}(\|\bs{s}-\bs{s}_0\|)\gtrsim 0.74$, and in conjunction with $\widehat{\beta} \approx 1$, indicates positive extremal dependence persists everywhere; this can be seen practically in Figure~\ref{fig:IrelandResults} with the estimates of $\chi_{0.95}(\bs{s}_1,\bs{s}_2) \gtrsim 0.3$. Estimates of $\psi$ and $L$ indicate stronger dependence in approximately a south-west to north-east direction.

\begin{table}[t!]
	\centering
	\caption{Parameter estimates and quantiles of the bootstrap distribution from 100 replicates of a stationary bootstrap with mean block size 10.}
	\vspace{10pt}
	\label{tab:IrelandResults}
	\begin{tabular}{rrrrrrrrrrr}
		\hline
		& $\kappa$ & $\lambda$ & $\beta$ & $\phi$ & $\nu$ & $\sigma$ & $\mu$ & $\delta$ & $\psi$ & $L$\\ 
		\hline
		Estimate	& 1.53 & 7.92 & 1.00 & 0.95 & 1.81 & 0.37 & $-0.22$ & 1.39  & -0.60 & 0.84\\ \hline
		5\% & 1.33 & 6.21 & 1.00 & 0.87  & 1.79 & 0.36 & $-0.29$ & 1.33 & -0.69 & 0.81\\ 
		95\% & 2.00 & 9.74 & 1.00 & 1.04 & 1.83 & 0.40  & $-0.14$ & 1.45 & -0.51 & 0.86\\ 
		\hline
	\end{tabular}
\end{table}

\begin{figure}[t!]
\centering
 \includegraphics[width=0.8\textwidth]{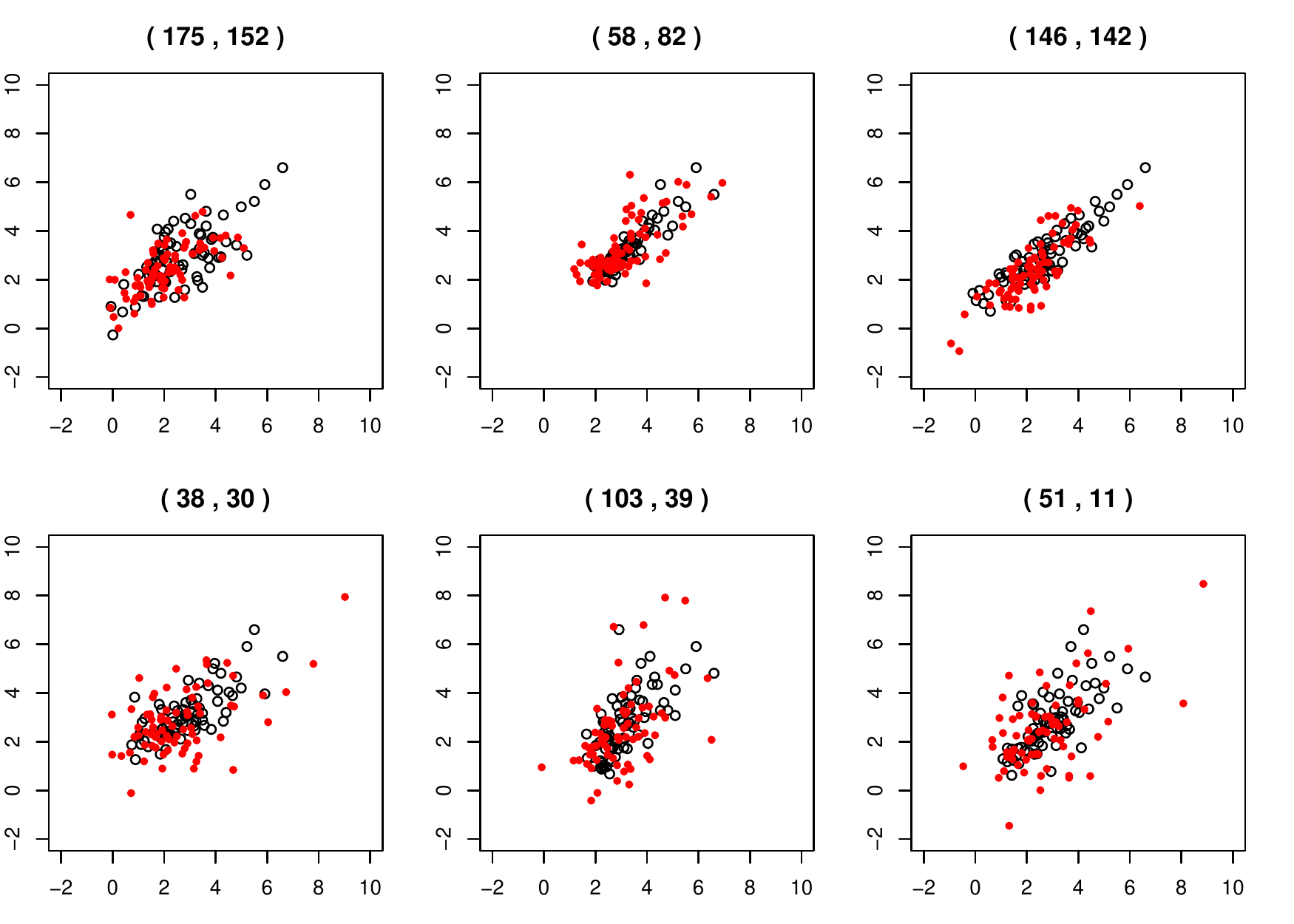}
  \caption{Pairwise plots of data (black) and simulations (red) conditioning upon the randomly-selected site 99 exceeding its 0.95 quantile. Pairs of sites are also randomly selected and displayed in the panel heading.}
 \label{fig:IrelandSimulations}
\end{figure}

Various possibilities for diagnostic plots are detailed in \citet{Wadsworth.Tawn:2019}. 
We focus here on checking that simulations from the fitted model are consistent with the data. Figure~\ref{fig:IrelandSimulations} displays example pairwise plots of data and simulations obtained conditioning on a randomly-selected site being large. Several more such plots can be viewed simply, and show that the data and simulations appear consistent. Furthermore, the right-hand panel of Figure~\ref{fig:IrelandResults} shows a model-based estimate of $\chi_{0.95}(\bs{s}_1,\bs{s}_2)$ and $\chi_{0.99}(\bs{s}_1,\bs{s}_2)$ overlaid on empirical estimates, using distances in the coordinate system transformed to account for anisotropy; the agreement appears satisfactory.

We use the fitted model to estimate the quantity $\Pr\{\max_{1 \leq j \leq D} X(\bs{s}_j) > v\}$ for large $v$, i.e., the probability that at least one location over the island exceeds a high threshold. This can be achieved using the approach described in \citet{Wadsworth.Tawn:2019} to estimate quantities of the form $\E\{g(X)\mid\max_{1 \leq j \leq D} X(\bs{s}_j)>v\}$ for any function $g$ using appropriately re-weighted simulations from the distribution of $X\mid X(\bs{s}_j)>v$. We construct the estimate by noting that
\begin{align*}
\Pr\left\{\max_{1 \leq j \leq D} X(\bs{s}_j) > v\right\} &= {\Pr\{ X(\bs{s}_i) > v\} \over \Pr\{X(\bs{s}_i)>v ~\mid~ \max_{1 \leq j \leq D} X(\bs{s}_j) > v\}},
\end{align*}
with the numerator having a known form due to Laplace margins and the denominator estimated by taking $g(X) = \mathbbm{1}(X(\bs{s}_j)>v)$ for any $j \in \{1,\ldots,D\}$. The estimate is displayed in Figure~\ref{fig:IrelandResults} for values of $v$ up to the 0.99995-quantile, along with the empirical estimates of this probability. There is good in-sample agreement, but the model permits extrapolation beyond the upper endpoint of the empirical distribution. The estimate of this distribution can be used to calculate suitable return levels, noting that the level $v_p$ defined by $\Pr\{\max_{1 \leq j \leq D} X(\bs{s}_j) > v_p\} = p$ represents the value exceeded once on average every $1/p$ summer days. An estimate of the 100-year return level of $\max_{1 \leq j \leq D} X(\bs{s}_j)$ is thus found by setting $p=1/(92 \times 100)$, and corresponds approximately to the 0.99998 quantile of the Laplace distribution. This value in turn is approximately the 543 year return level for $X(\bs{s}_j)$, i.e., the value of the temperature at any given site.

\section{Conclusion}\label{sec:conclusion}
Modeling spatial extremes relies on assumptions about the joint tail decay rate. These assumptions are especially crucial when the ultimate goal is to extrapolate beyond historical data in order to estimate spatial risk measures. For mathematical elegance and because of their asymptotic characterization, max-stable and Pareto processes have played a key role and are frequently used in practice. However, when the data display asymptotic independence or a weakening strength of extremal dependence at increasing levels, such asymptotic models are no longer appropriate. In this paper, we have reviewed recently proposed alternative spatial models that provide increased flexibility to capture the sub-asymptotic behavior, with a limiting extremal dependence structure that can be precisely characterized. We have mainly focused on random scale (or location) mixture constructions and the conditional spatial extremes model of \citet{Wadsworth.Tawn:2019}, which allow to bridge asymptotic dependence and independence regimes in a single parametrization, and we have also briefly mentioned other related approaches such as max-mixture or max-id models. While most of these models are constructed from underlying common random factors that affect the overall dependence structure of the process, preventing them from capturing complete independence at large distances, the conditional extremes approach circumvents this limitation at the price of being more heavily parametrized and more difficult to interpret ``unconditionally''. Further research is needed to develop relatively parsimonious spatial models that combine a flexible tail dependence structure in terms of tail and spatial decay rates, an intuitive unconditional interpretation, and feasible inference in large dimensions.

From a computational perspective, the ``sub-asymptotic'' dependence models presented here are generally somewhat easier to handle in high dimensions than their asymptotic counterparts (e.g., max-stable models), yet progress still needs to be made to use them on really big data. {In this paper, we have focused on likelihood-based inference, and the \texttt{R} code used to fit the models considered in the real data applications in Section~\ref{sec:applications} can be obtained upon request from the authors.} While flexible random scale constructions of the form \eqref{eq:Huseretal} or the \citet{Huser.Wadsworth:2019} model fitted by censored likelihood inference can be applied up to dimensions about $D=30$--$50$, the conditional spatial extremes model of \citet{Wadsworth.Tawn:2019} is currently limited to dimensions of the order $D=300$--$500$ (without censoring). Computational speed-up may be obtained for censored likelihood approaches by exploiting pseudo-Monte Carlo methods or hierarchical matrix decompositions for the calculation of multivariate Gaussian or Student's $t$ distributions \citep{deFondeville.Davison:2018,Genton.etal:2018,Cao.etal:2019,Beranger.etal:2020}; 
or even  by using proper scoring rules instead of maximum likelihood as in \citet{deFondeville.Davison:2018}. However, to tackle problems in truly higher dimensions, sparse models with a fundamentally different probabilistic structure \citep{Engelke.Hitz:2020,Engelke.Ivanovs:2021} need to be devised. While \citet{Engelke.Hitz:2020} developed sparse multivariate Pareto distributions, which allow factorization of densities on graphs, this framework does not apply to the asymptotic independence or hybrid models presented in this paper. 

Finally, most applications in the extreme-value literature assume that the data come from a stationary and isotropic process. This is usually valid in small regions, but with modern high-dimensional data, complex statistical models are often necessary to capture spatio-temporal non-stationarities. In our environmental applications, we have shown how to incorporate geometric anisotropy, but further research is required to develop (potentially semi-parametric) spatial extremes models with flexible joint tail structures that are valid over large domains. 

\section*{Funding Information}
This publication is based upon work supported by the King Abdullah University of Science and Technology (KAUST) Office of Sponsored Research (OSR) under Award No. OSR-CRG2017-3434.

\section*{Acknowledgments}
We thank Thomas Opitz for helpful discussions.





\bibliographystyle{CUP}
\bibliography{BigBib}

\end{document}